\documentclass[useAMS,usenatbib]{mn2e}
\usepackage{txfonts}
\usepackage{graphicx}
\usepackage{natbib}
\usepackage{amssymb}
\usepackage{color}

\newcommand\lax{\>\vcenter{\hbox{$<$\hskip-.75em\lower1.0ex\hbox{$\sim$}}}\>}
\newcommand\uax{\>\vcenter{\hbox{$>$\hskip-.75em\lower1.0ex\hbox{$\sim$}}}\>}

\newcommand{\aap}{A{\&}A}

\newcommand{\apj}{ApJ}
\newcommand{\apjl}{ApJL}

\newcommand{\pra}{Phys. Rev. A}

\title[Ionization and dissociation equilibrium in strongly-magnetized helium atmosphere]{Ionization and dissociation equilibrium in strongly-magnetized helium atmosphere}
\author[K. Mori and J.S. Heyl]{Kaya Mori$^{1, 2}$\thanks{e-mail: kaya@cita.utoronto.ca} and
  Jeremy S. Heyl$^{3}$\thanks{e-mail: heyl@physics.ubc.ca}\\
$^{1}$ Department of Astronomy and Astrophysics,
  University of Toronto, 50 St. George Street, Toronto, Ontario, M5S
  3H4, Canada\\ 
$^{2}$ Canadian Institute for Theoretical Astrophysics, University of Toronto, 60 St. George Street, Toronto, Ontario, M5S 3H8, Canada\\
$^{3}$ Department of Physics and Astronomy; 
University of British Columbia; Vancouver, BC V6T 1Z1, Canada; 
Canada Research Chair}

\begin{document}
\pagerange{\pageref{firstpage}--\pageref{lastpage}} \pubyear{2006}

\maketitle 

\label{firstpage}

\begin{abstract}
  Recent observations and theoretical investigations of neutron stars
  indicate that their atmospheres consist not of hydrogen or iron but
  possibly other elements such as helium.  We calculate the
  ionization and dissociation equilibrium of helium in the
  conditions found in the atmospheres of magnetized neutron stars.
  For the first time this investigation includes 
  the internal degrees of freedom of the helium molecule.  We found
  that at the temperatures and densities of neutron star atmospheres
  the rotovibrational excitations of helium molecules are populated.
  Including these excitations increases the expected abundance of
  molecules by up to two orders of magnitude relative to calculations
  that ignore the internal states of the molecule; therefore, if the
  atmospheres of neutron stars indeed consist of helium, helium
  molecules and possibly polymers will make the bulk of the atmosphere
  and leave signatures on the observed spectra from neutron stars.
  We applied our calculation to nearby radio-quiet neutron stars with
  $B_{dipole}\sim10^{13}$--$10^{14}$ G.  
  If helium comprises their atmospheres, our study indicates that
  isolated neutron stars with $T_{BB}\sim10^6$ K such as
  RXJ0720.4-3125 and RXJ1605.3+3249 will have He$^+$ ions
  predominantly, while isolated neutron stars with lower temperature ($T_{BB}\sim5\times10^5$
  K) such as RXJ1856.5-3754 and RXJ0420.0-5022 will have
  some fraction of helium molecules.     
  We found that ionization, dissociation and electric excitation
  energies of helium molecules are larger than 100 eV at 
  $B\uax10^{13}$ G. On the other hand, rotovibrational excitation
  energies are in the 
 range of 10--100 eV at $B=10^{12}$--$10^{14}$ G. If
  helium molecules are abundant, their spectroscopic signatures may be
  detected in the optical, UV and X-ray band.  
   
\end{abstract}



\section{Introduction} 

Hydrogen has been considered as the surface composition of isolated
neutron stars (INSs) because gravitational stratification forces the
lightest element to the top of the atmosphere \citep{alcock80}. Only a
tiny amount of material is required to constitute an optically thick
layer on the surface \citep{romani87}.  However, recent studies of
\citet{chang04_1} and \citet{chang04_2} have shown that the NS surface
may be composed of helium or heavier elements since hydrogen may be
quickly depleted by diffusive nuclear burning.  Observationally,
helium and heavier element atmospheres have been proposed for
interpreting the spectral features observed in several INS partially
because the existing hydrogen atmosphere models do not reproduce the
observed spectra \citep{sanwal02, hailey02, vankerkwijk06}.  However,
atomic and molecular data in the strong magnetic field regime are
scarce for non-hydrogenic elements.  Accurate atomic and molecular
data are available mostly for the He$^+$ ion \citep{pavlov05}, the
helium atom \citep{neuhauser87, demeur94}, He$_2^{3+}$
\citep{turbiner04_2} and He$_2^{2+}$ 
\citep{turbiner06_3}. Helium molecular binding energies have been
crudely calculated by density functional theory \citep{medin06_1,
  medin06_2} (hereafter ML06).  Unlike hydrogen atmospheres
\citep{lai97}, the ionization and dissociation balance in
strongly-magnetized helium atmosphere has not been investigated yet.

In this paper, we extend our Hartree-Fock type calculation
\citep{mori02} to helium molecules in the Born-Oppenheimer
approximation. For molecular ions that exist in strong magnetic fields 
($B=10^{12}$--$10^{14}$ G), we achieved  $\lax$1\% and $\lax$10\%
agreement in binding energies and vibrational energies in comparison
with other more accurate studies mainly on hydrogen
molecules. Including numerous electronic, vibrational and rotational
states, we studied ionization and dissociation equilibrium in helium atmospheres at
$B=10^{12}$--$10^{14}$ G. We also applied our calculations to several
INSs which may have helium atmospheres on their surfaces.


\section{Molecular binding and vibrational energy}
\label{sec_bind}

At first we adopt the Born-Oppenheimer approximation and neglected any
effects associated with motion of atoms and molecules in a magnetic
field. Later, we will discuss rotovibrational states (\S\ref{sec_rovib}) and how the finite nuclear mass modifies
results (\S\ref{sec_finite_mass}).  In the Landau regime ($\beta_Z>1$
where $\beta_Z \equiv B/(4.7\times10^9Z^2~\mbox{G})$ and $Z$ is the
atomic number), bound electrons in an atom and molecule
are well specified by two quantum numbers $(m,\nu)$. $m$ is the 
absolute value of a magnetic 
quantum number (which is negative to lower the 
total energy in strong magnetic fields) and $\nu$ is a longitudinal quantum number along the
field line. We consider only tightly-bound states with 
$\nu=0$. Electronic excited states with $\nu>0$ have small binding 
energies, therefore their population in the atmosphere is tiny due to 
small Boltzmann factors and pressure ionization.  Hereafter we denote 
atomic and molecular energy states as $(m_1, m_2, m_3, m_4, ...)$.  

\begin{figure} 
\resizebox{\hsize}{!}{\includegraphics{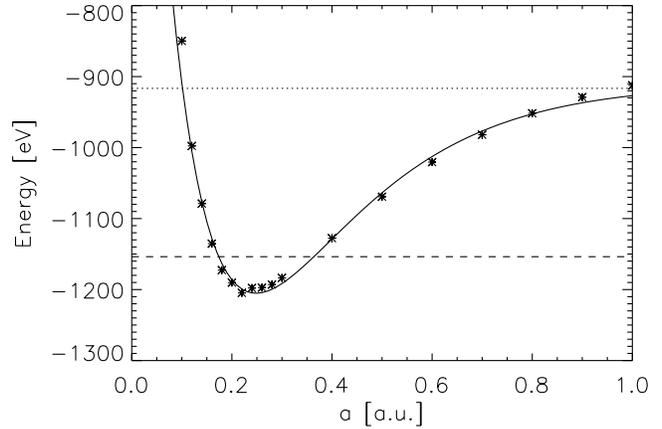}}
\caption{Binding energy curve of He$_2$ at $B=10^{12}$ G. The solid
  curve is the fitted Morse function. The dashed line corresponds to
  the energy of two Helium atoms in the ground state $(0,1)$
  ($=-E_m+D_m$). The dotted line corresponds to the summed energy of one
  Helium atom in $(0,3)$ state and the other in $(1,2)$ state
  ($=-E_m+\tilde{D}_m$). \label{fig_he2_energy}}
\end{figure}

We computed molecular binding energies with a simple modification to 
our Hartree-Fock type calculation for atoms \citep{mori02}. We 
replaced the nuclear Coulomb term $V_m(z)$ in the Schr{\"o}dinger equation by \citep{lai92} 
\begin{equation}
\tilde{V}_m(z)=V_m\left(z-\frac{a}{2}\right)+V_m\left(z+\frac{a}{2}\right)
\end{equation} 
where
\begin{equation}
V_m(z)\equiv \int d^2\vec{r}_\perp \frac{|\Phi_m(\vec{r}_\perp)|^2}{r} .
\end{equation}
The function $\Phi_{m}$ is the ground state Landau wavefunction, and
$a$ is the separation between two nuclei. We added $\frac{Z^2e^2}{a}$
as the Coulomb repulsion energy between the two nuclei. We computed
binding energies with a grid size $\Delta a\sim0.1$ [a.u.] up to
$a\sim1$ [a.u.] and $\Delta a\sim0.01$ [a.u.] near the energy
minimum. Figure \ref{fig_he2_energy} shows the binding energy curve of
He$_2$ at $B=10^{12}$ G fitted with the Morse function defined as
\citep{morse29}
\begin{equation}
U(a) = \tilde{D}_m \{1-\exp{\left[-\zeta(a-a_0)\right]}\}^2-E_m
\label{eq_morse}
\end{equation}
where $E_m (>0)$ is the molecular binding energy for an electronic
state $m$ (in this paper this usually denotes the magnetic quantum
number of the outermost electron), $a_0$ is the separation between two
nuclei at the minimum energy. We defined two different dissociation
energies: $D_m \equiv E_m-2E_{a}(gs)$ and $\tilde{D}_m \equiv
E_m-E_{a}(m_a, m_b)-E_{a}(m_c, m_d)$. $E_{a}(gs)$ is the ground state
energy of an atom (e.g., $(0,1)$ state for Helium atom) and
$E_{a}(m_a, m_b)$ is the energy of an atom in the $(m_a, m_b)$
state. Each of the atomic $m$ quantum numbers ($m_a, m_b, m_c$ and
$m_d$) corresponds to one of the molecular $m$ quantum numbers ($m_1,
m_2, m_3$ and $m_4$) so that $E_{a}(m_a, m_b)+E_{a}(m_c, m_d)$ is the
smallest. For instance, helium atoms in $(0,3)$ and $(1,2)$ state are
the least bound system into which He$_2$ in the ground state
(i.e. $(0,1,2,3)$ state) will dissociate. Note that a molecule
dissociates to atoms and ions when $E_m < 2E_a(gs)$, while the
molecular binding energy approaches $E_{a}(m_a,m_b)+E_{a}(m_c, m_d)$
at large $a$.

The calculated binding energy values are not smooth near the energy
minimum (Figure \ref{fig_he2_energy}). This is due to our numerical
errors.  Binding energy does not change by more than 0.1\% for $\Delta
a =0.01$ [a.u.] near the energy minimum.  We determined $E_m$ from the
fitting procedure using the function given by equation 
(\ref{eq_morse}) since we found that it provides more accurate results
than the minimum energies from our grid calculation. However, in most cases, $E_m$ from our grid calculation and
the fitted $E_m$ do not differ by more than 1\%. We computed $D_m$ and
$\tilde{D}_m$ using the atomic data we calculated numerically.

Our results for H$^+_2$ and H$_2$ are in good agreement with
\citet{turbiner03} (hereafter TL03) and \citet{lai96} (LS96) with less
than 1\% deviation in total binding energy
(Tab.~\ref{tab_h_bind_energy}). TL03 performed highly accurate
variational studies mainly on one-electron molecular systems (e.g.,
H$^+_2$, H$^{2+}_3$, He$^{3+}_2$). LS96 studied hydrogen molecular
structure similarly by a Hartree-Fock calculation in the adiabatic
approximation. While our calculation takes into account higher Landau
levels using perturbation theory, the difference in binding energies
by including higher Landau levels is tiny for helium atoms and
molecules at $B\ge10^{12}$ G.

Similar to hydrogen, the ground state configuration is $(m_1, m_2)=(0,
1)$ for He$^{2+}_2$ and $(m_1, m_2, m_3, m_4)=(0,1,2,3)$ for He$_2$ at
$B\ge10^{12}$ G. The accurate comparisons for helium molecules are
\citet{demeur94} and ML06 who computed He$_2$ binding energy by
Hartree-Fock theory (table \ref{tab_he_bind_energy}). Our results
agree with ML06 within 1\%. ML06 also computed helium molecular
binding energies using density functional theory (ML06)
\footnote{Although the results of ML06 are mostly from density
  functional calculation, they showed some results from Hartree-Fock
  calculation based on \citet{lai96} for comparison. }. However,
their DFT results are less accurate than those of Hartree-Fock
calculation; the binding energies are overestimated by $\sim10$\%
(ML06).

\begin{table}  
\caption{Total binding energy [eV] of H$^+_2$ (left) and H$_2$
  (right). \label{tab_h_bind_energy}}
\begin{center}
\begin{tabular}{|c|c|c|c|c||c|c|c|}
\hline
$B_{12}$ & $m$ & This work & LS96 & TL03 & $m_1, m_2$ & This work & LS96\\
\hline
0.1 & 0  & 102 &  99.9 & 102  & 0, 1  & 162  & 161  \\
\hline
1  &  0 & 233 & 232 & 233 &  0, 1 & 370 & 369\\
   &  1 & 162 &  162 &  162 &  0, 2 & 336  & 337\\  
\hline 
10 &  0 & 484 & 486  & 486 &  0, 1 & 772 & 769  \\
   &  1 & 356 & 356  & 356 &  0, 2 & 713  & 709 \\ 
\hline
\end{tabular}
\begin{quote} 
 {\scshape Notes:} LS96: \citet{lai96}, TL03: \citet{turbiner03}
\end{quote}
\end{center}
\end{table}

\begin{table}   
\caption{Total binding energy [eV] of He$_2$ in comparison with ML06. \label{tab_he_bind_energy}}
\begin{center}
\begin{tabular}{|c|c|c|c|c|}
\hline
$B_{12}$ & This work & ML06\\
\hline
1 & 1207  & 1202 \\
10  & 2733  & 2728 \\
100 & 5597  & 5598 \\
\hline
\end{tabular}
\end{center}
\end{table}


\subsection{Electronic excitation} 

The electronic excited states in molecules occupy higher $(m,\nu)$
states than those in atoms.  Since excitation energies from the ground
state to $(0, 1, 2, m)$ states with $m>3$ are small, there may be
numerous tightly-bound electronic excited states until they dissociate
into atoms and ions at large $m$. We did not consider excited states
with $\nu>0$ because their binding energies are small therefore they
are likely to be dissociated. 

\begin{figure} 
\resizebox{\hsize}{!}{\includegraphics{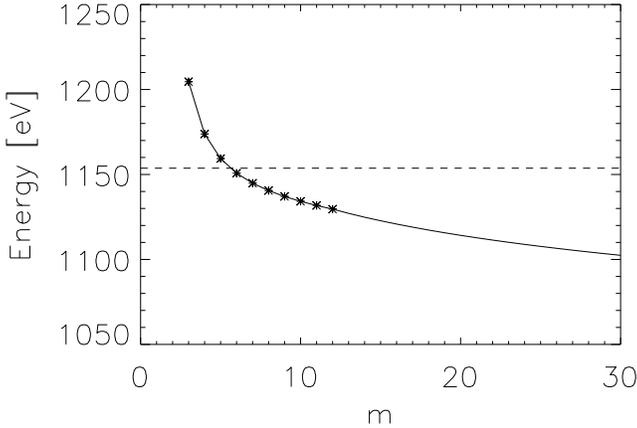}}
\caption{Binding energy curve of He$_2$ in $(0,1,2,m)$ states at
  $B=10^{12}$ G. The asterisk points are the binding energies from our
  numerical calculations. The dashed line corresponds to the energy of
  two Helium atoms in the ground state. \label{fig_he2_excite}}
\end{figure}

We calculated binding energies for the
$(0,1,2,m)$ state up to $m=9$ and estimated binding energies for
higher $m$ states using the well-known $m$ dependence of the energy
spacing $\Delta E_m \sim \ln\left(\frac{2m+3}{2m+1}\right)$
\citep{lai92}. We found that the difference between the exact solutions 
and those from the scaling law is tiny at $m\ga9$. Figure \ref{fig_he2_excite} shows the He$_2$ binding
energy of $(0, 1, 2, m)$ states at $B=10^{12}$ G. Note that the
excited states of He$_2$, $(0,1,2,m)$ with $m>5$, are unbound with
respect to two atoms in the ground state at $B=10^{12}$ G.

\section{Rotovibrational excitation} 
\label{sec_rovib} 
 
We consider molecular excitation levels associated with vibrational
and rotational motion of molecules in a magnetic field. In contrast to
the field-free case, the strong magnetic field induces molecular
oscillations with respect to the field line similar to a two-dimensional
harmonic oscillator. Accordingly, there are three types of molecular
motion; vibration along and transverse to the magnetic field and
rotation around the magnetic field. Hereafter we briefly
describe energy levels of rotovibrational states. 

Strictly speaking, the aligned and transverse vibrations are coupled \citep{khersonskii84,
  lai96}. However, using perturbation theory,
\citet{khersonskii85} has shown that the coupling energy is tiny
(less than 1\%) compared to the total binding energy. Neglecting the 
coupling, the rotovibrational energy levels are well approximated by 
\begin{equation} 
\epsilon_{N\Lambda V}=\epsilon_V+\epsilon_{N\Lambda}. 
\end{equation}
$\epsilon_V$ is the aligned vibrational energy given by 
\begin{equation} 
\epsilon_V=\hbar\omega_{\parallel}\left(V+\frac{1}{2}\right)-\frac{(\hbar\omega_{\parallel})^2}{4\tilde{D}_m}\left(V+\frac{1}{2}\right)^2. 
\label{eq_align}
\end{equation}
The integer $V (\ge0)$ is the quantum number for the aligned vibration and 
$\hbar\omega_{\parallel}$ is the aligned vibrational energy quanta
\citep{morse29, khersonskii85}. 

On the other hand, the transverse rotovibration energy 
($\epsilon_{N\Lambda}$) consists of transverse vibration
and rotation around the magnetic field axis and it is given as
\citep{khersonskii85}   
\begin{equation} 
\epsilon_{N\Lambda}=\hbar\omega_{t}\left(N+\frac{|\Lambda|+1}{2}\right)-\frac{\Lambda}{2}\hbar\Omega_B.  
\label{eq_trans}
\end{equation} 
The integer $N (\ge0)$ is the quantum number for the transverse
vibration, while the integer $\Lambda$ is the projection of
angular momentum in the B-field direction \citep{khersonskii85}. 
$\hbar\Omega_B$ is the nuclear cyclotron energy ($=Z \hbar
eB/(Am_{p}c)=6.3(Z/A)B_{12}$ [eV] where $A$ is the atomic mass) and
$\omega_t=\sqrt{4\omega_{\perp}^2+\Omega_B^2}$. $\omega_{\perp}$ is
the transverse vibrational energy quanta. The nuclear cyclotron
energy term takes into account the magnetic restoring force on the
nuclei \citep{lai96}. In the following 
subsections, we calculate vibrational energy quanta in the
Born-Oppenheimer approximation. 


\subsection{Aligned vibrational excitation} 
\label{sec_align}

In the Born-Oppenheimer approximation, the motion of two nuclei along
the magnetic field is governed by the binding energy curves determined
in section \ref{sec_bind}. It is therefore straightforward to
calculate aligned vibrational energy quanta using the results from
section \ref{sec_bind}. We fit the Morse function given by equation (\ref{eq_morse}) to molecular binding energies as a function
of the nuclear separation $a$. Once $\zeta$ is
determined, the aligned vibrational energy quanta is given by (LS96)
\begin{equation} 
\hbar\omega_{\parallel}=\hbar\zeta\left(\frac{2\tilde{D}_m}{\mu}\right)^{1/2} 
\end{equation} 
where $\mu$ is the reduced mass of the two nuclei in units of the
electron mass (918 for H$_2$ and 3675 for He$_2$).  At large $a$,
another electron configuration is mixed with tightly-bound states
\citep{lai92}. Since configuration interaction is neglected in our
calculation, our $\zeta$ values (therefore $\hbar\omega_{\parallel}$)
are overestimated by 10--30\% in comparison with LS96 (table
\ref{tab_vib_h}).  We found that $\zeta$ is nearly identical for
different (tightly-bound) electronic excited states. Therefore we
computed $\hbar\omega_{\parallel}$ for electronic excited states with
large $m$ (for which we did not perform grid calculation) using
$\zeta$ from the lower excited states. In most cases, our aligned
vibrational energy quanta agree with other more accurate results within $\sim10$\%
(table \ref{tab_vib_h}). Table \ref{tab_vib_he2+} compares our results
for He$_2^{2+}$ with \citet{turbiner06_3} at various magnetic fields. Table \ref{tab_vib_he} shows aligned
vibrational energy quanta for helium molecular ions along with some
results on He$_2^{3+}$ from \citet{turbiner04_2}. 

\begin{table}  
\caption{Aligned vibrational energy quanta $\hbar\omega_{\parallel}$
  [eV] of  H$^+_2$ (left) and H$_2$ (right). \label{tab_vib_h} }
\begin{center}
\begin{tabular}{|c|c|c|c||c|c|}
\hline
$B_{12}$ & This work & LS96 & TL03 & This work & LS96\\
\hline
0.1 & 3.2 & 2.0 &  2.4 & 3.3 & 3.0\\
0.5 & 6.1 & 4.9 & -- & 6.4 & 7.2  \\
1  &  7.2 & 6.6 & 7.5  & 11 & 9.8 \\ 
2 & 12 & 9.0  & -- & 14 &  13 \\
5 & 14 & 13 & -- & 17 & 19 \\
10 & 17 & 17 & 20 & 29 & 25\\
\hline
\end{tabular}
\begin{quote} 
 {\scshape Notes:} We multiplied the results of TL03 by a factor of two to correct the
discrepancy due to different definitions of aligned vibrational energy
quanta. 
\end{quote}
\end{center}
\end{table}

\begin{table}  
\caption{Comparison of aligned vibrational energy quanta $\hbar\omega_{\parallel}$
  [eV] of  He$_2^{2+}$ with \citet{turbiner06_3}. \label{tab_vib_he2+} }
\begin{center}
\begin{tabular}{|c|c|c|}
\hline
$B_{12}$ & This work & TG06\\
\hline
0.0235 & 1.2 & 0.82\\
0.235 & 2.4 & 3.3\\
2.35  &  10 & 11 \\ 
23.5 & 34 & 32   \\
\hline
\end{tabular}
\begin{quote} 
 {\scshape Notes:} Similar to table \ref{tab_vib_h}, we multiplied the results of TG06 by a factor of two. 
\end{quote}
\end{center}
\end{table}

\begin{table}  
\caption{Aligned vibrational energy quanta $\hbar\omega_{\parallel}$
  [eV] of  helium molecular ions. \label{tab_vib_he}}
\begin{center}
\begin{tabular}{|c|c|c|c|c|}
\hline
$B_{12}$ & He$_2^{3+}$ & He$_2^{2+}$  & He$_2^{+}$ & He$_2$  \\
\hline
1 & 4.3 (2.7) & 5.9  &  7.8 &  10 \\
10  & 7.5 (9.8) & 22 &  30 &  32 \\ 
100 & 13 &  64 &  78 &  81 \\
\hline
\end{tabular}
\begin{quote}
 {\scshape Notes:} The numbers in the brackets show the results from \citet{turbiner04_2}. 
\end{quote}
\end{center}
\end{table}

It is apparent that the discrepancies with other calculations are
larger for helium molecules than for hydrogen molecules. There are two effects both of which
reduce the accuracy of our results particularly for highly-ionized
molecules at low magnetic fields. First, we did not take into account
configuration interaction, while the work by Turbiner et al. employed
the full 2-dimensional 
variation energy calculation. The degree of configuration interaction
is larger at small $\beta_Z$ since the increasing effects of the
nuclear Coulomb field mix different electron 
configurations. Second, highly-ionized molecular ions are either
unbound or weakly-bound at low magnetic fields, therefore the numerical
errors in binding energies significantly affect determination of
$\zeta$ since the binding energy curve is shallow. Nevertheless, the
accuracy of vibrational energies of highly-ionized molecules such as
He$_2^{3+}$ is irrelevant for dissociation balance since they are
unbound or their abundance is negligible (section \ref{sec_id}). For the
abundant molecular ions in $10^{12}$--$10^{14}$ G (e.g., He$_2^{+}$
and He$_2$), we expect the accuracy of aligned vibrational energy is $\lax10$\%. 


\subsection{Transverse vibrational excitation} 
\label{sec_trans}
  
We calculated the energy curve as a function of transverse position of
nuclei $R$ following Ansatz A described in section IIIB of
LS96. We fixed $a$ to the equilibrium separation $a_0$ and supposed that the
two nuclei are located at $(\pm R/2, \pm a_0/2)$. As LS96 pointed out,
this method is appropriate for small $R$ ($\la \hat{\rho}$ where
$\hat{\rho}$ is the cyclotron radius) and gives only an upper limit to
transverse vibrational energy quanta $\hbar\omega_{\perp}$.  We
replaced the nuclear Coulomb term in the Schr\"odinger equation $V(z)$
by
\begin{equation} 
V(z,R/2)=V_{m}\left(z-\frac{a_0}{2},\frac{R}{2}\right)+V_{m}\left(z+\frac{a_0}{2},\frac{R}{2}\right)
\end{equation}  
where 
\begin{equation} 
V_{m}(z, R/2)=\int {d^2 \vec{r}_\perp 
  \frac{|\Phi_{m}(\vec{r}_\perp)|^2}{|\vec{r}_\perp-\vec{R}/2|}}.  
\end{equation} 
We added $\frac{Z^2e^2}{(a_0^2+R^2)^{1/2}}$ as the Coulomb repulsion
energy between two nuclei.  Once we calculated the molecular binding
energy at different $R$ grid points, we fit a parabolic form
$\frac{1}{2}\mu\omega_{\perp}^2R^2$ to binding energies at $R \la
\hat{\rho}$. We found that $\hbar\omega_{\perp}$ is
nearly identical for different electronic excitation levels.
Therefore, we adopted $\hbar\omega_{\perp}$ of the ground state for
electronic excited states. Our transverse vibrational energy quanta
agree with those of LS96 and TL03 within 10\% (table
\ref{tab_h_trans}). Table \ref{tab_he_trans} shows the results for
helium molecular ions in comparison with those for He$_2^{3+}$ from
\citet{turbiner04_2}. Compared to aligned vibrational energy, our results for transverse vibrational
energy are in better agreement with other more accurate results. The
better agreement is well-understood because the transverse potential
well is deeper than in the aligned direction in magnetic field, 
therefore our results are less subject to the numerical errors as
discussed in section \ref{sec_align}. 

\begin{table}  
\caption{Transverse vibrational energy quanta $\hbar\omega_{\perp}$
  [eV] of H$^+_2$ (left) and H$_2$ (right).  The numbers in the brackets are transverse
  vibrational energy [eV] computed using the perturbation
  theory. \label{tab_h_trans}}
\begin{center}
\begin{tabular}{|c|c|c|c||c|c|}
\hline
$B_{12}$ & This work & LS96 & TL03 & This work & LS96\\
\hline
0.1 & 2.8 (3.3) & 3.1 & 2.9  & 2.5 (3.2) & 2.6 \\
0.5 & 8.7 (9.5) & 9.8 & -- & 8.7 (9.1) & 8.7  \\
1  & 14 (15) & 16 & 15  & 14 (15) & 14 \\ 
2 & 22 (24) &  25  & -- & 22 (22) &  23 \\
5 & 42 (41) & 45 & --  & 41 (40) & 42\\
10 & 63 (63) & 70 & 66  & 64 (61) & 65\\
\hline
\end{tabular}
\end{center}
\end{table}

\begin{table}  
\caption{Transverse vibrational energy quanta [eV] $\hbar\omega_{\perp}$ of
  helium molecular ions. The numbers in the brackets are transverse
  vibrational energy [eV] computed using the perturbation theory. \label{tab_he_trans}}
\begin{center}
\begin{tabular}{|c|c|c|c|c|}
\hline
$B_{12}$ & He$_2^{3+}$ & He$_2^{2+}$  & He$_2^{+}$ & He$_2$  \\
\hline
1 &  9.8 (10) $^{\mbox{(a)}}$ &  10 (11) &  9.7 (10) &  10 (9.6) \\
10  & 46 (60) $^{\mbox{(a)}}$ &  50 (49) &  48 (45) &  50 (43) \\ 
100 & 205 (193) &  216 (203) & 217 (195) & 216 (182) \\
\hline
\end{tabular}
\begin{quote}
(a) Transverse vibrational energy quanta of He$_2^{3+}$ ions from
\citet{turbiner04_2} are 11 eV and 51 eV at $B_{12}=1$ and 10
respectively.    
\end{quote} 
\end{center}
\end{table} 

\subsubsection{Perturbative approach}
 \label{sec_perturb}

It is also possible to estimate the transverse vibrational energy
perturbatively, by calculating the lowest order perturbation to the
energy of the molecule induced by tilting it.  We assume that
the energy of the tilted molecule is almost the same as the energy
required to displacing the electron cloud relative to the molecule by
an electric field ($E$) :
\begin{equation}
\epsilon_\kappa^{(1)} = \left < \kappa \left | eE x \right | \kappa \right > = 0.
\label{eq:1}
\end{equation}
The first-order change to the wavefunction is 
\begin{equation}
\kappa^{(1)} = \sum_{\kappa'} \frac{\left < \kappa' \left | eE x \right | \kappa
  \right >}{\epsilon_{\kappa'}^{(0)} - \epsilon_\kappa^{(0)}} \kappa'
\label{eq:2}
\end{equation}
where $\epsilon_\kappa^{(0)}$ is the unperturbed energy of the state $\kappa$ and $\kappa'$
denotes the other states of the system.

The expectation value of $x$ for this situation is
\begin{equation}
r \sin \theta = \langle x \rangle = eE \sum_{\kappa'} \frac{\left | \left < \kappa' \left
 |  x \right | \kappa
  \right >\right |^2}{\epsilon_{\kappa'}^{(0)} - \epsilon_\kappa^{(0)}}
\label{eq:3}
\end{equation}
where $r$ is half the distance between the nuclei.  We can solve for
the value of $eE$ that we need to apply to give a particular
displacement and substitute it into the expression for the energy of
the state to second order
\begin{equation}
\epsilon_\kappa^{(2)} = \left(  e E \right )^2 \sum_{\kappa'} 
\frac{\left | \left < \kappa' \left | x \right | \kappa  \right >\right |^2}{\epsilon_{\kappa'}^{(0)} - \epsilon_{\kappa}^{(0)}} = r^2 \sin^2 \theta \left ( \sum_{\kappa'} \frac{\left | \left < \kappa' \left |  x \right | \kappa  \right >\right |^2}{\epsilon_{\kappa'}^{(0)} - \epsilon_\kappa^{(0)}} \right )^{-1}
\label{eq:4}
\end{equation}
The only states that contribute to the sum have $m'=m\pm 1$ for the
electronic states of the molecule (the rotational states do not count
because that is what we are examining).

States that have been excited along the field ($\nu>0$) or by
increasing the Landau number do not have
much overlap in the integral in the numerator and also a large energy
difference in the denominator.

The frequency for low amplitude oscillations is given by
\begin{equation}
\omega^2_\perp = \frac{2 r^2}{I} \left ( \sum_{\kappa'} \frac{\left | \left < \kappa' \left |  
x \right | \kappa
  \right >\right |^2}{\epsilon_{\kappa'}^{(0)} - \epsilon_\kappa^{(0)}} \right )^{-1} =   \frac{1}{M}  \left ( \sum_{\kappa'} \frac{\left | \left < \kappa' \left |  
x \right | \kappa
  \right >\right |^2}{\epsilon_{\kappa'}^{(0)} - \epsilon_\kappa^{(0)}} \right )^{-1}
\label{eq:6}
\end{equation}
where $I$ is the moment of inertia of the molecule, $2 M r^2$ ($M$ is
the nuclear mass). The size of the molecule has cancelled out.   

For a single electron system in the ground state if we assume that the
bulk of the contribution to the sum in the equation is given by the
$m=1,\nu=0$  state we have
\begin{equation}
\omega_\perp \approx \frac{\sqrt{2}}{{\hat \rho}} \left ( \frac{\epsilon_{m=1,\nu=0} -
    \epsilon_{m=0,\nu=0}}{M} \right
)^{1/2} \left | \left < f_{m=1,\nu=0} |  f_{m=0,\nu=0} \right >\right |
\label{eq:7}
\end{equation} 
where the final term is the overlap of the longitudinal wavefunction of the
two states.  

For a multielectron system, evaluating equation (\ref{eq:4}) and
(\ref{eq:6}) is somewhat
more complicated.  For clarity of nomenclature we shall write the
wavefunctions of the various electronic states of the molecule as ${\cal K}'$.
The symbol ${\cal K}$ denotes the state that we are focused upon.
The change in the energy of the system due to an applied electric field is
\begin{equation}
\epsilon^{(1)} = \left < {\cal K} \left | N e E {\bar x} \right | {\cal K} \right > = 0
\label{eq:8}
\end{equation}
where $N$ is the number of electrons and
\begin{equation}
N \bar x = \sum_j x_j
\label{eq:9}
\end{equation}
where $j$ counts over the electrons in the molecule. We have assumed
that the multielectron wavefunctions are normalized such that 
$\left <  {\cal K} | {\cal K} \right >=1$

The first-order change to the wavefunction is 
\begin{equation}
{\cal K}^{(1)} = \sum_{\cal K'} \frac{\left < {\cal K}' \left | NeE \bar x \right | \cal K
  \right >}{\epsilon_{\cal K'}^{(0)} - \epsilon_{\cal K}^{(0)}} \cal K'
\label{eq:10}
\end{equation}
where $\epsilon_{\cal K'}^{(0)}$ is the unperturbed energy of the state $\cal K'$.

Now let us calculate the expectation value of $\bar x$ for this situation,
\begin{equation}
r \sin \theta = \langle \bar x \rangle = N e E \sum_{\cal K'} \frac{\left | \left < {\cal K'} \left | \bar x \right | {\cal K}
  \right >\right |^2}{\epsilon_{\cal K'}^{(0)} - \epsilon_{\cal K}^{(0)}}
\label{eq:11}
\end{equation}
where $r$ is half the distance between the nuclei.  We can solve for
the value of $NeE$ that we need to apply to give a particular
displacement and substitute it into the expression for the energy of
the state to second order
\begin{equation}
\epsilon_{\cal K}^{(2)} = \left( NeE \right )^2 
 \sum_{\cal K'} \frac{\left | \left < {\cal K'} \left | \bar x \right | {\cal K}
  \right >\right |^2}{\epsilon_{\cal K'}^{(0)} - \epsilon_{\cal K}^{(0)}}
 = r^2 \sin^2 \theta \left (  \sum_{\cal K'} \frac{\left | \left < {\cal K'} \left | \bar x \right | {\cal K}
  \right >\right |^2}{\epsilon_{\cal K'}^{(0)} - \epsilon_{\cal K}^{(0)}}
\right )^{-1}
\label{eq:12}
\end{equation}

In strongly magnetized atoms or molecules, it is
natural to expand the wavefunctions in terms of the ground Landau
level using various values of $m$.  We assume that the wavefunction 
for each electron is written as
\begin{equation}
\kappa_j = \frac{1}{\sqrt{2\pi}} \Phi_{m_j}(\rho) f_j(z)  e^{im_j\phi}.
\label{eq:13}
\end{equation}
In this case we have
\begin{equation}
\left < \kappa_i | x | \kappa_j \right > = \frac{{\hat \rho}}{2} \sqrt{ 2 ( m_j + 1 )} \left < f_i | f_j \right > ~{\rm if}~ m_i = m_j + 1
\label{eq:14}
\end{equation}
otherwise it vanishes.

Combining these results yields an estimate for the frequency of low
amplitude oscillations of
\begin{equation}
\omega_\perp \approx \frac{N \sqrt{2}}{{\hat \rho}} \left (
\frac{\epsilon_{m+1} - \epsilon_{m}}{M (m+1)} \right )^{1/2} \left | \left < f_i |  f \right >\right |
\label{eq:15}
\end{equation} 
where the subscript on energy labels the value of $m$ in the
outermost shell.  The number of electrons appears in this equation
because the expectation value of $\bar x$ is a factor $N$ smaller than
the expectation value of $x$ for the single shifted electron.

For the ground state of the molecule we have $m+1=N$ yielding a
simpler expression,
\begin{equation}
\omega_\perp \approx \frac{\sqrt{2}}{{\hat \rho}} \left ( N \frac{\epsilon_{m+1} -
    \epsilon_{m}}{M} \right )^{1/2} \left | \left < f_i |  f \right >\right |.
\label{eq:16}
\end{equation} 
Table \ref{tab_h_trans} and \ref{tab_he_trans} compare the perturbative estimates with the
numerical calculations.  Within the approximations made the agreement
is encouraging.


\section{Rotovibrational spectrum} 
\label{sec_rovib_spec}

\begin{figure} 
\resizebox{\hsize}{!}{\includegraphics{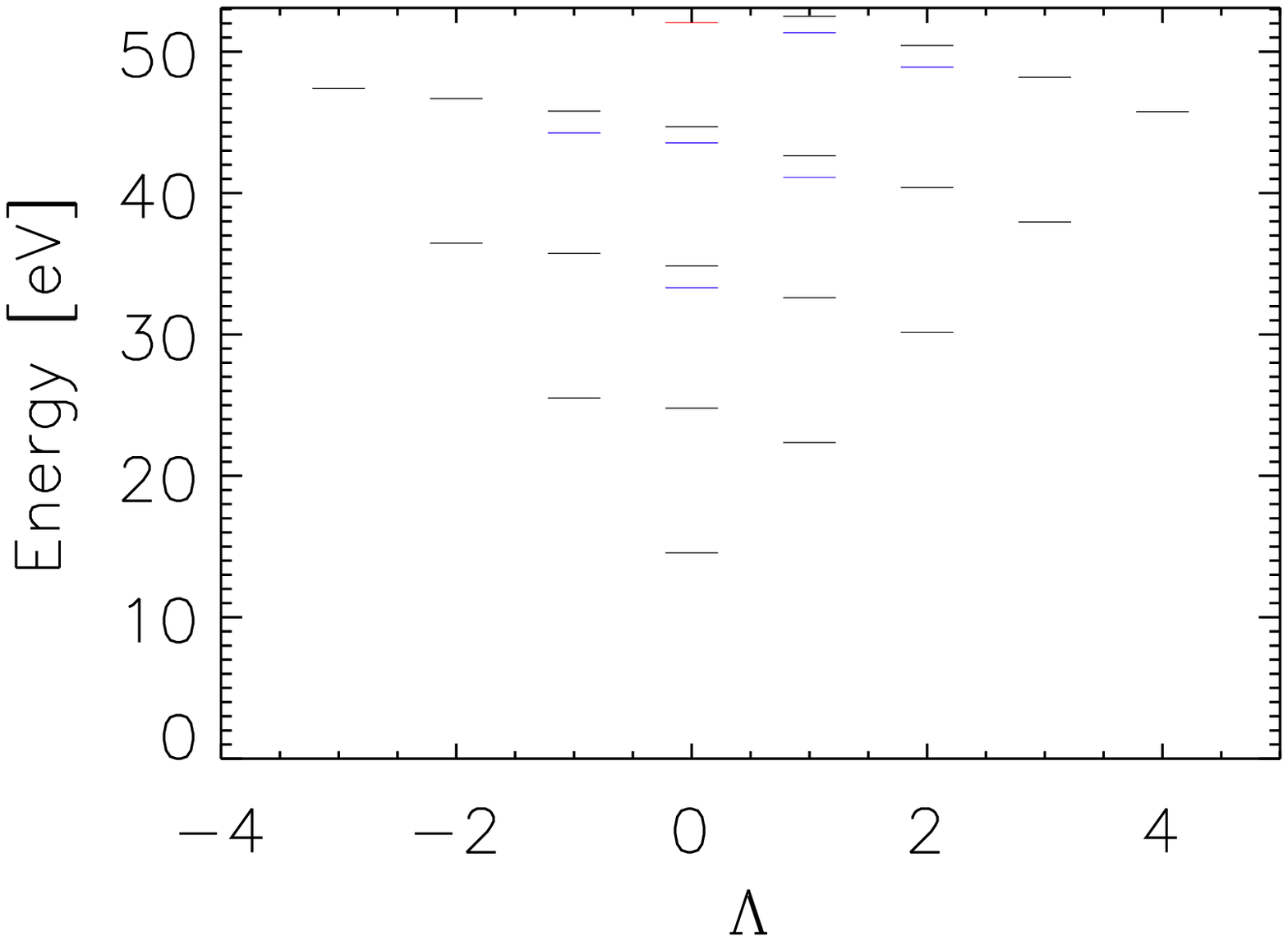}}
\resizebox{\hsize}{!}{\includegraphics{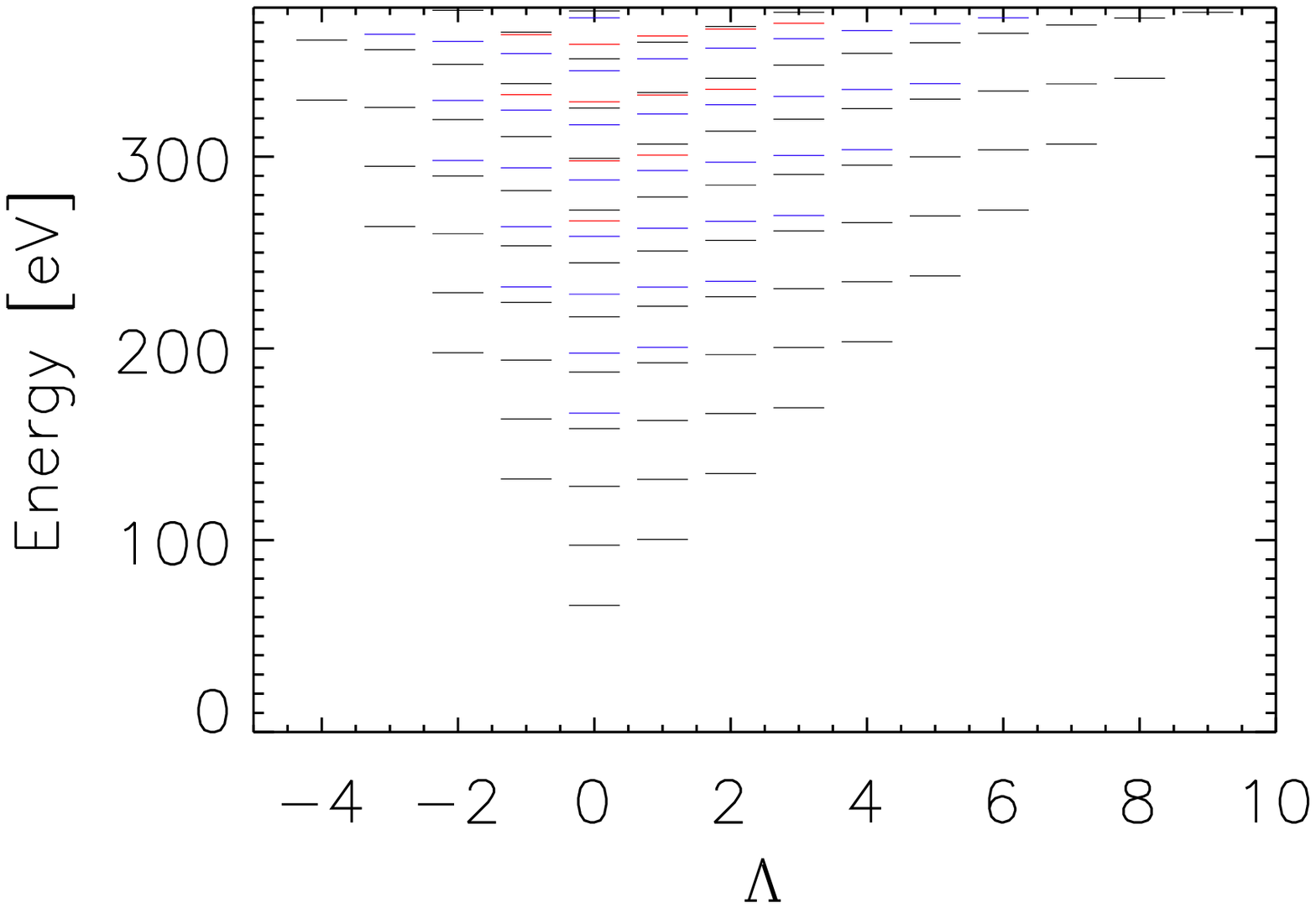}}
\caption{Rotovibrational energy spectrum ($\epsilon_{N\Lambda V}$) of
  He$_2$ in the ground state at $B=10^{12}$ G (top) and $10^{13}$ G
  (bottom). The zero energy corresponds to $E=-E_m$, while the
  uppermost horizontal line corresponds to $E=-E_m+D_m$ above which
  the molecular states become dissociated. The black, blue and red
  lines indicate $N=0, 1$ and 2 state. For each $N$ and $\Lambda$, the
  horizontal lines indicate energy levels for different $V$
  states. \label{fig_he2_rovib}}
\end{figure}

Given the aligned and transverse vibrational energies, we construct
the rotovibrational spectra of helium molecules.  
From the equations (\ref{eq_align}) and (\ref{eq_trans}), the molecular system has a finite zero-point
energy associated with aligned and transverse vibration
\begin{equation}
\epsilon_{000}=\frac{1}{2}(\hbar\omega_t + \hbar\omega_{\parallel})
\end{equation}  
when $N=\Lambda=V=0$. Therefore, the actual dissociation energy will
be reduced from $D_m$ by $\epsilon_{000}$ (LS96).  Figure
\ref{fig_he2_rovib} shows the rotovibrational energy spectrum of He$_2$
in the ground state at $B=10^{12}$ G and $10^{13}$ G
respectively. Table \ref{tab_rovib} shows the number of rotovibrational
states of various helium molecular ions. The number of rotovibrational
states decreases for higher electronic excited states.  The magnetic
field dependence is more complicated for the following reasons.
The number of aligned vibrational levels generally increases with $B$.
On the other hand, the number of transverse vibrational ($N$) and
rotational ($\Lambda$) levels increases with $B$ at $B\lax B_Q$
($B_Q=4.414\times10^{13}$~G) while it decreases with $B$ at $B\uax B_Q$. This is because the ion cyclotron
energy ($\propto B$) dominates over $\hbar\omega_\perp$ ($\propto
B^{1/2}$) at $B\uax B_Q$. It should be noted that only those rotovibrational
states with excitation energies that are smaller than or similar to
the thermal energy have large statistical weights in the partition
function (\S\ref{sec_id}). 

\begin{table}  
\caption{Number of rotovibrational states in helium molecular ions. \label{tab_rovib}}
\begin{center}
\begin{tabular}{|c|c|c|c|c|}
\hline
 & $(m_1, m_2, m_3, m_4)$ & $B_{12}=1$ & $B_{12}=10$ & $B_{12}=100$  \\
\hline
He$_2^{3+}$  &  (0) & 0  & 0  &  1 \\
\hline
He$_2^{2+}$  &  (0,1) & 0  &  8  &  12 \\ 
             &  (0,2) & 0  &  0 &   1 \\ 
\hline 
He$_2^{+}$  & (0,1,2)  & 98  &  177 &  87 \\
 & (0,1,3) & 5 & 79 & 56 \\ 
 & (0,1,4) & 0 & 46 & 42 \\
 & (0,1,5) & 0 & 31 & 35 \\
\hline
He$_2$  & (0,1,2,3)  & 27  &  132 &  87 \\
 & (0,1,2,4)  & 2  &  77  &  61 \\
 & (0,1,2,5)  & 0  &  57  &  52 \\
 & (0,1,2,6)  & 0  &  50  &  46 \\
\hline
\end{tabular}
\end{center}
\end{table} 

\section{Effects of finite nuclear mass}
 \label{sec_finite_mass}

The separation of the center-of-mass motion is non-trivial when a
magnetic field is present \citep{herold81}. The total
pseudomomentum ($\vec{K}$) is often used to take into account motional
effects in a magnetic field since $\vec{K}$ is a constant of motion \citep{lai01}. 
In the following subsections, we discuss two effects associated with finite nuclear mass in
strong magnetic fields. We denote the binding energy in the assumption
of fixed nuclear location (e.g., Born-Oppenheimer approximation) as $\epsilon^{(0)}_\kappa (<0)$ for an
electronic state $\kappa$. Since we consider states with $n=0$ and
$\nu=0$, the relevant quantum numbers are $\kappa=\{m_i\}$ ($i$
denotes each bound electron in multi-electron atoms and molecules).

\subsection{Finite nuclear mass correction}
 \label{sec_nuc_mass}

The assumption of zero transverse pseudomomentum introduces an additional term $s_\kappa\hbar \Omega_B$ in the binding energy 
\citep{herold81, wunner81}. $\Omega_B$ is the nuclear cyclotron energy and
$s_\kappa=\sum_i m_i$ is the sum of magnetic quantum numbers for a
given electronic state $\kappa$ (e.g., $s_\kappa=4$ for He$_2$
molecule in the ground state). However, the scheme assuming the zero transverse pseudomomentum does not
necessarily give the lowest binding energies at $B\uax B_Q$.  Instead,
LS95 and LS96 estimated lower binding energies at $B\uax B_Q$ using another
scheme which relaxed the assumption of the zero transverse
pseudomomentum. A more rigorous calculation was performed for He$^+$ ion
by \citet{bezchastnov98} and \citet{pavlov05}. An application of such
schemes to multi-electron systems is beyond the scope of this paper.
We will discuss the limitation of our models at very high magnetic
field in \S\ref{sec_id}.

\subsection{Motional Stark effects}
 \label{sec_ms}

When an atom or molecule moves across the magnetic field
$\vec{B}$, a motional Stark electric field
$\vec{E}_{MS}=\frac{\vec{K}\times\vec{B}}{MC}$ is induced in the
center-of-mass frame. $M$ is the mass of atom or molecule. The
Hamiltonian for the motional Stark field is given by 
\begin{equation} 
H_{MS}=e\frac{\vec{K}\times\vec{B}}{Mc}\vec{r}=\Omega_{B0} K_\perp x
\end{equation}
where $\Omega_{B0}\equiv \frac{eB}{Mc}$. 
For a given pseudomomentum $\vec{K}$, the motional Stark field
separates the guiding center of the nucleus and that of the electron
by  
\begin{equation} 
R_K=\frac{c|\vec{K}\times\vec{B}|}{eB^2}=\frac{c K_\perp}{eB}. 
\end{equation}

Since the motional Stark field breaks the cylindrical symmetry
preserved in magnetic field, it is non-trivial to evaluate motional
Stark field effects. A non-perturbative (therefore more rigorous)
approach has been applied only for one-electron systems
\citep{vincke92, potekhin94, pavlov05}. However, such an approach is
quite complicated and time-consuming especially for multi-electron
atoms and molecules. Therefore, following LS95, we considered two
limiting cases: (1) $R_K \ll \hat{\rho}$ and (2) $R_K \gg \hat{\rho}$
and determined general formula which can be applied to a wide range of
$B$ and $K_\perp$. For diatomic molecules, we can apply a nearly
identical scheme used for calculating transverse vibrational energy to
the both cases.

\subsubsection{Centered states}
\label{sec_center}

When the energy shift caused by motional Stark field is smaller than
the spacing between binding energies, the perturbation approach is
applicable (\cite{pavlov93} and \S~\ref{sec_perturb}). The first order perturbation energy
$\langle\kappa|H_{MS}|\kappa\rangle$ vanishes since the matrix element
$\langle\kappa |x|\kappa\rangle=0$. The second order perturbation energy
is given by
\begin{equation}
\epsilon^{(2)}_\kappa=\sum_{\kappa'} \frac{|\langle
  \kappa'|H_{MS}|\kappa\rangle|^2}{\epsilon^{(0)}_{\kappa'}-\epsilon^{(0)}_\kappa}=\frac{K^2_\perp}{2M}\alpha^{(2)}_\kappa 
\end{equation}
Among the various electronic states $\kappa'=\{n'm'\nu'\}$, only
$n'=0, m'=m\pm1, \nu'=0$ state have
non-negligible contribution to $\epsilon^{(2)}_\kappa$ since the other
states have large $\epsilon^{(0)}_{\kappa'}-\epsilon^{(0)}_\kappa$
and/or vanishing matrix element
$\langle\kappa'|x|\kappa\rangle$. Since the overlap integral of
longitudinal wavefunction is close to unity \citep{pavlov93},
$\alpha^{(2)}_\kappa$ is given by
\begin{equation}
\alpha^{(2)}_\kappa\simeq\hbar\Omega_{B0}\left(\frac{m+1}{\epsilon_m-\epsilon_{m+1}+\hbar\Omega_{Bi}}+\frac{m}{\epsilon_m-\epsilon_{m-1}-\hbar\Omega_{Bi}}\right) 
\end{equation}
where $\epsilon_m=-\epsilon^{(0)}_\kappa (> 0)$ where $m$ denotes the
magnetic quantum number of the outermost electron of electronic state
$\kappa$. The 2nd term is zero for the ground state. 
Following \citet{pavlov93}, we define the anisotropic mass
$M_{\perp,\kappa}$ as 
\begin{equation} 
M_{\perp, \kappa}\equiv \frac{M}{1-\alpha^{(2)}_\kappa} > M.  
\end{equation}
The transverse energy characterized by $K_\perp$ is given by
\begin{equation} 
E_{\perp, \kappa}=\frac{{K_\perp}^2}{2M_{\perp, \kappa}}.
\label{eq_perp} 
\end{equation} 

The perturbation method is valid when
$|\epsilon^{(2)}_\kappa|\ll|\Delta\epsilon^{(0)}_\kappa|$
\citep{pavlov93} where $K_{\perp,\kappa}$ is given by
\begin{equation}
K_{\perp,\kappa} = \left(\frac{2M|\Delta\epsilon^{(0)}_\kappa|}{\alpha^{(2)}_\kappa}\right)^{1/2}. 
\end{equation} 
where $|\Delta\epsilon^{(0)}_\kappa|$ is the spacing of the zeroth order
energies (typically $|\epsilon_m-\epsilon_{m+1}|$). 

\subsubsection{Decentered states}
 \label{sec_decenter}

When $R_K \gg \hat{\rho}$, it is convenient to utilize the so-called
decentered formalism (LS95). We replace the nuclear Coulomb
term by  
\begin{equation}
V_{m}(z, R_K)= \int {d^2 \vec{r}_\perp 
  \frac{|\Phi_{m}(\vec{r}_\perp)|^2}{|\vec{r}_\perp+\vec{R}_K|}}, 
\end{equation}
and compute binding energies at different $R_K$ grid points. 
For diatomic molecules, we replace the nuclear Coulomb term by 
\begin{equation}
V(z, R_K)=V_m\left(z-\frac{a_0}{2}, R_K\right)+V_m\left(z+\frac{a_0}{2}, R_K\right).  
\end{equation}
The grid calculation for $R_K$ is identical to the one for transverse
vibrational energy in \S\ref{sec_trans} except that the Coulomb
repulsion term between two nuclei is $\frac{Z^2e^2}{a_0}$ (instead of
$\frac{Z^2e^2}{(a^2_0+R^2_K)^{1/2}}$) since motional Stark field
shifts the guiding center of the two nuclei by $R_K$ in the transverse
direction but the separation between the two nuclei is still $a_0$.

\begin{figure} 
\resizebox{\hsize}{!}{\includegraphics{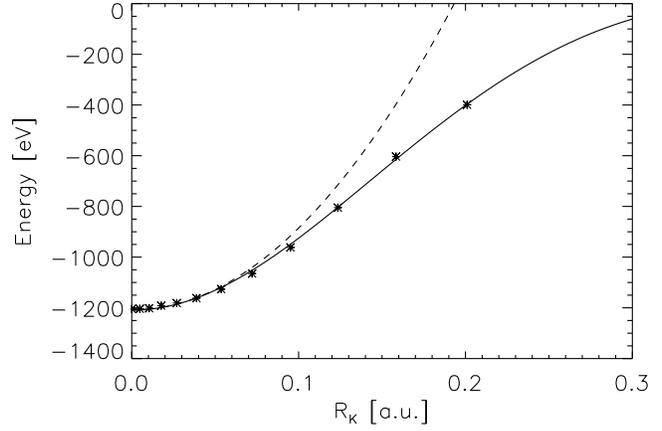}}
\caption{The transverse energy of He$_2$ molecule at $B=10^{12}$~G. 
  The asterisks are the binding energies from our numerical 
  calculation and they are fitted with the function given in (\ref{eq_fit}). The dashed line shows the energy curve from the
  perturbation method discussed in \S\ref{sec_perturb} and \S\ref{sec_center}. 
  \label{fig_ms_fit}}
\end{figure}

LS95 found that the binding energy curves are well fit by the
following formula.
\begin{equation}
E_\perp(K_\perp)+\epsilon_\kappa^{(0)}=-A_1\left(\ln\frac{1}{A_2+A_3
  R^2_K}\right)^2
 \label{eq_fit} 
\end{equation}
where $A_1, A_2$ and $A_3$ are the fit parameters.
$E_\perp(R_K)=E_\perp(K_\perp)$ is the transverse energy and
$\epsilon_\kappa^{(0)}$ is the binding energy in the infinite nuclear
mass approximation. Figure \ref{fig_ms_fit} shows the binding energy
curve of He$_2$ molecule as a function of $R_K$ at $B=10^{12}$ G. At
small $R_K$, the fitted function is well matched with the results from
the perturbation approach. Although mixing between different $m$
states is ignored, a comparison with \citet{potekhin98-2} and
\citet{potekhin94} for hydrogen atoms indicates this approach gives
better than 30\% accuracy over a large range of $K_\perp$
\citep{lai95}. This is adequate for our purpose of investigating
ionization and dissociation balance.

Along with the finite nuclear mass term discussed in
\S\ref{sec_nuc_mass}, the electronic energy of an atom or molecule
moving with transverse pseudomomentum $K_\perp$ is given by 
\begin{equation} 
\epsilon_{\kappa}(K_\perp)=\epsilon^{(0)}_{\kappa}+s_{\kappa}\hbar\Omega_B+E_\perp(K_\perp). \label{eq_bind_energy}
\end{equation} 
Note that $\epsilon^{(0)}_{\kappa}<0$ and both the second and third term 
decrease the binding energy. Later, we will discuss the validity of this
approach at $B\uax B_{Q}$. 


\section{Ionization and dissociation equilibrium}
 \label{sec_id}

We have investigated the ionization and dissociation balance of magnetized
helium atmospheres including the following chemical reaction
channels. Table \ref{tab_ion_e} and \ref{tab_dis_e} list ionization
and dissociation energies of various helium ions and molecular
ions in the assumption with fixed nuclear location. We did not take into account the He$^{-}$ ion since its
ionization energy is $\la20$ [eV] and He$^{-}$ is not abundant at all
in the temperature range considered here ($T\ga10^5$ K).
\begin{itemize} 
\item Ionization 
\begin{itemize} 
\item[(1)]{He$^+ \leftrightarrow \alpha$ + e}
\item[(2)]{He $\leftrightarrow$ He$^{+}$ + e} 
\item[(3)]{He$^{2+}_2$ $\leftrightarrow$ He$^{3+}_2$ + e} 
\item[(4)]{He$^{+}_2$ $\leftrightarrow$He$^{2+}_2$ + e}
\item[(5)]{He$_2$ $\leftrightarrow$He$^{+}_2$ + e}
\end{itemize} 
\item Dissociation 
\begin{itemize} 
\item[(6)]{He$^{3+}_2$ $\leftrightarrow$  $\alpha$ + He$^{+}$} 
\item[(7)]{He$^{2+}_2$ $\leftrightarrow$  $\alpha$ + He} 
\item[(8)]{He$^{2+}_2$ $\leftrightarrow$  He$^{+}$ + He$^+$} 
\item[(9)]{He$^{+}_2$ $\leftrightarrow$ He$^{+}$ + He} 
\item[(10)]{He$_2$ $\leftrightarrow$ He + He} 
\end{itemize} 
\end{itemize} 

\begin{table}  
  \caption{Ionization energy [eV] of helium atom and molecular ions. \label{tab_ion_e}}
  \begin{center}
\begin{tabular}{|c|c|c|c|c|c|c|}
\hline
$B_{12}$ & He$^+$ & He & He$_2^{3+}$ & He$_2^{2+}$  & He$_2^{+}$ & He$_2$  \\
\hline
1 &  418 &  159  &  391 & 418 &  261 &  137 \\
10  & 847  & 331 &  885 &  947  &  603 &  298 \\ 
100 & 1563 & 629 &  1833 &  1912 & 1209  & 643 \\
\hline
\end{tabular}
\end{center}
\end{table} 

\begin{table}  
\caption{Dissociation energy $D_m$ [eV] of helium molecular ions. The
  numbers in the brackets are $\tilde{D}_m$ [eV]. The columns without a number
  indicate that there is no bound molecular state with respect to
  ions and atoms. The zero-point energy correction is not included. \label{tab_dis_e}}
\begin{center}
\begin{tabular}{|c|c|c|c|c|}
\hline
$B_{12}$ & He$_2^{3+}$ & He$_2^{2+}$  & He$_2^{+}$ & He$_2$  \\
\hline
1 & -- (--) &  -- (97.8) & 74.6 (251)  & 53.1 (290) \\
10  & 37.8 (37.8) &  137 (359) & 410 (725)  & 378 (806)  \\ 
100 & 270 (270)  & 619 (973)  & 1198 (1708) & 1212 (1915) \\
\hline
\end{tabular}
{\scshape Notes:} Recent work by \citet{turbiner04_2} and
\citet{turbiner06_3} shows He$_2^{3+}$ and He$_2^{2+}$ ion can exist
at $B\ga(2.4-2.6)\times10^{12}$ G. Our results are roughly consistent
with theirs.  
\end{center}
\end{table} 

The Saha-Boltzmann equation for ionization balance 
is given as, 
\begin{eqnarray}
\frac{n_i}{n_{i+1}n_e}&=&\frac{z_i}{z_{i+1}z_e}
\end{eqnarray} 
where $z_e$ is the partition function for an electron, 
\begin{equation} 
z_e=2\left(\frac{m_ekT}{2\pi\hbar^2}\right)^{3/2}\frac{\eta_e}{\tanh{\eta_e}},  
\end{equation} 
$\eta_e=\hbar\omega_B/2kT$ and $\omega_B=eB/m_ec$ is the electron
cyclotron frequency \citep{gnedin74}.  The quantity $z_i$ is the
partition function for ionization state $i$,
\begin{equation}
z_i=\left(\frac{M_i kT}{2\pi\hbar^2}\right)^{1/2}
    \frac{\eta_i}{\sinh{\eta_i}} \int_0^{\infty} \frac{K_\perp d K_\perp}{2\pi\hbar^2}
    \sum_{\kappa} w_{i,\kappa}(K_\perp) \exp{\left(-\frac{\epsilon_{i,
    \kappa}(K_\perp)}{kT}\right)}
\end{equation} 
where $E_i$, $Z_i$ and $M_i$ are the ground state energy, the charge
and the mass of an ion $i$. $\eta_i = \hbar\Omega_{Bi}/2kT$ where
$\Omega_{Bi} = Z_ieB/M_ic$ is the ion cyclotron
frequency. $\epsilon_{i,\kappa}(K_\perp) (< 0)$ is the binding energy
of a bound state $\kappa$ given in equation (\ref{eq_bind_energy}).
$w_{i, \kappa}$ is the occupation probability. In general, $w_{i,
  \kappa}(K_\perp)$ is a function of $Z_i$, $Z_p$ (the charge of
perturbing ions, usually the effective charge of the plasma),
$\epsilon_{i,\kappa}(K_\perp)$ and $\rho$ (the plasma density). $w$
depends not only on the type of ion $i$, the 
electronic state $\kappa$ and the transverse
pseudomomentum $K_\perp$, but also it is different for each bound
electron. Obviously, electrons in the outer shells are subject to
a stronger electric field from neighboring ions than those in the inner
shells. We explicitly computed $w_{i,\kappa}$ for all the bound
electrons using the electronic microfield distribution of
\citet{potekhin02}.

The Saha-Boltzmann equation for dissociation balance is given as 
\begin{eqnarray}
\frac{n_j}{n_{k}n_{l}}&=&\frac{\tilde{z}_j}{z_{k}z_{l}} 
\end{eqnarray} 
where $\tilde{z}_j$ is the partition function for molecular ionization
state $j$,
\begin{eqnarray}
\tilde{z}_{j}&=&\displaystyle \left(\frac{M_j kT}{2\pi\hbar^2}\right)^{1/2}
\frac{\eta_j}{\sinh{\eta_j}} \int_0^{\infty} \frac{K_\perp d
  K_\perp}{2\pi\hbar^2} \sum_{\kappa}
w_{j,\kappa}(K_\perp)\nonumber\\
& & \times \exp{\left(-\frac{\epsilon_{j,\kappa}(K_\perp)}{kT}\right)} \sum_{N,  
  \Lambda, V}^{\epsilon_{N\Lambda V}<D_\kappa}\exp{\left(-\frac{\epsilon_{N\Lambda V}}{kT}\right)}
\end{eqnarray} 
where $E_j$, $Z_j$ and $M_j$ are the ground state energy, the charge
and the mass of a molecular ion $j$. $\eta_j= \hbar\Omega_{Bj}/2kT$
where $\Omega_{Bj} = Z_jeB/M_jc$ is the molecular ion cyclotron
frequency. $\epsilon_{j, \kappa}(K_\perp) (<0)$ and
$w_{j,\kappa}(K_\perp)$ are the binding energy and occupation
probability of an electronic bound state $\kappa$. $\epsilon_{N\Lambda
  M} (>0)$ is the excitation energy of a rotovibrational state
$(N,\Lambda, V)$. We took the summation $\sum_{N,\Lambda, V}$ until
$\epsilon_{N\Lambda V}$ exceeds the dissociation energy $D_\kappa$.
The set of the Saha-Boltzmann equations along with the condition for
the baryon number conservation and charge neutrality are
iteratively solved until we reached sufficient convergence in $\Delta
n_e/n_e$ ($< 10^{-6}$) where $n_e$ is the density of free
electrons. The convergence was achieved rapidly in most cases (less
than 10 iterations were required).

\begin{figure} 
\resizebox{\hsize}{!}{\includegraphics{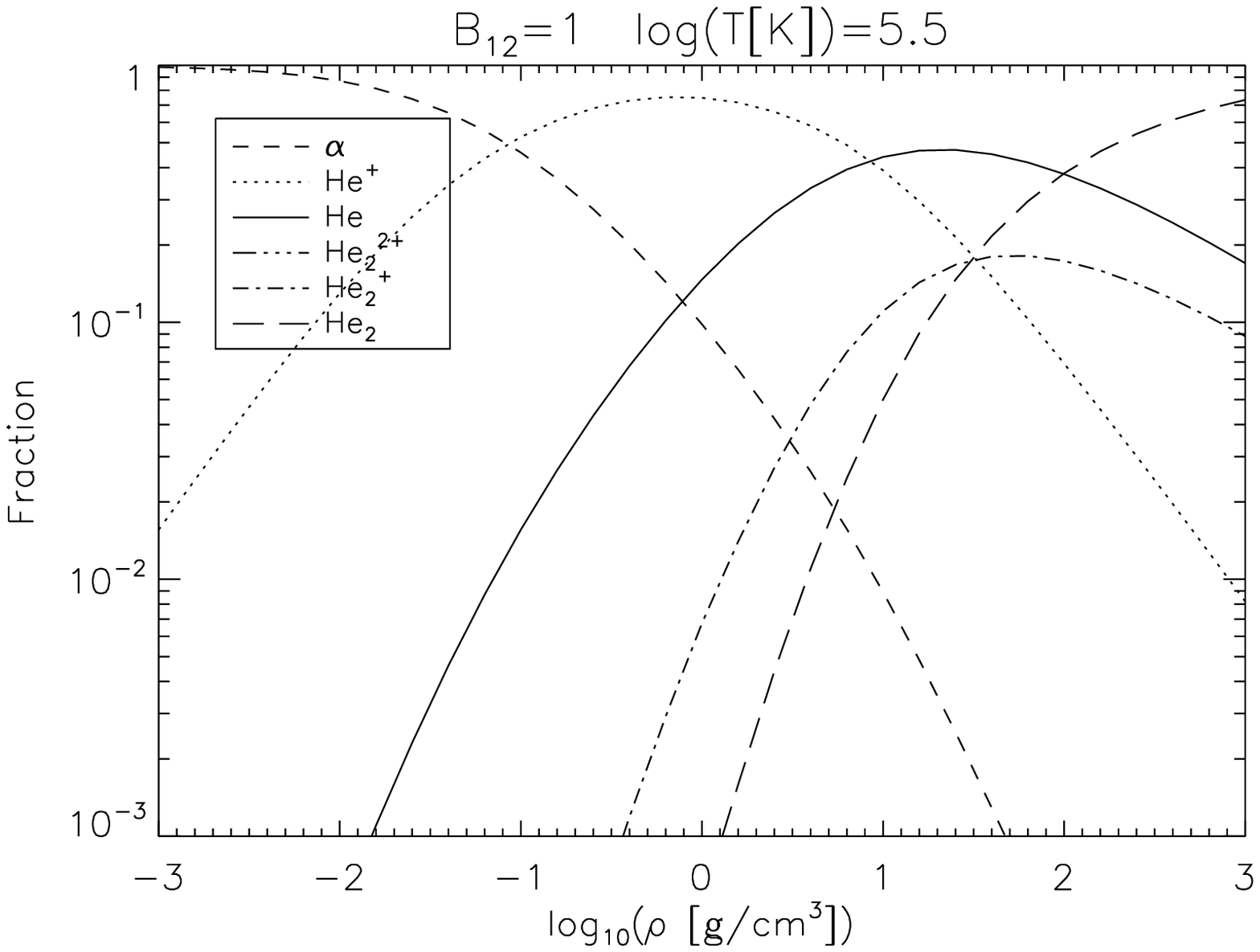}}
\resizebox{\hsize}{!}{\includegraphics{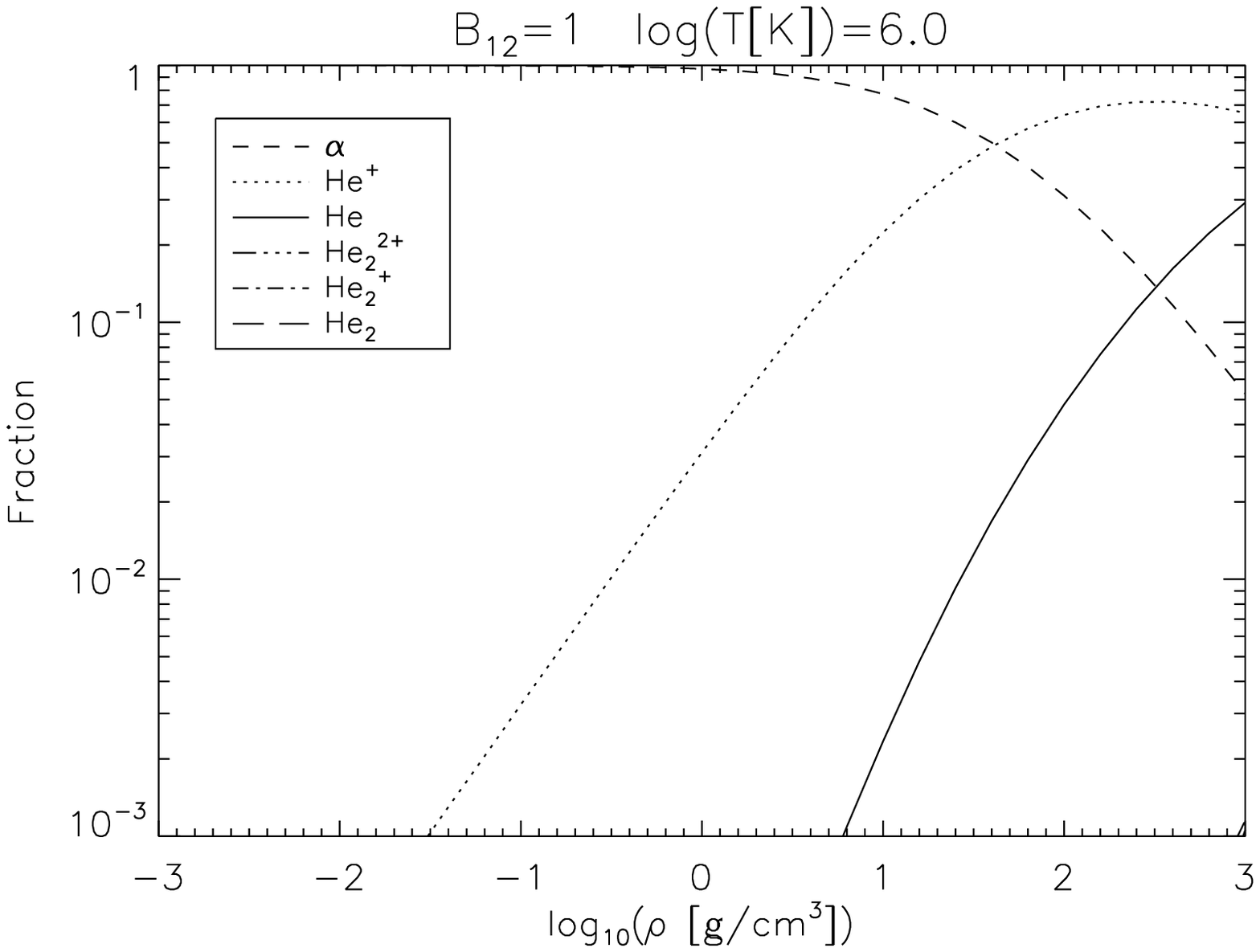}}
\caption{Ionization and dissociation balance of helium at  $T=10^{5.5}$ K (top) and $10^6$ K (bottom) at $B=10^{12}$ G.
  \label{fig_he12_rho}}
\end{figure}

\begin{figure} 
\resizebox{\hsize}{!}{\includegraphics{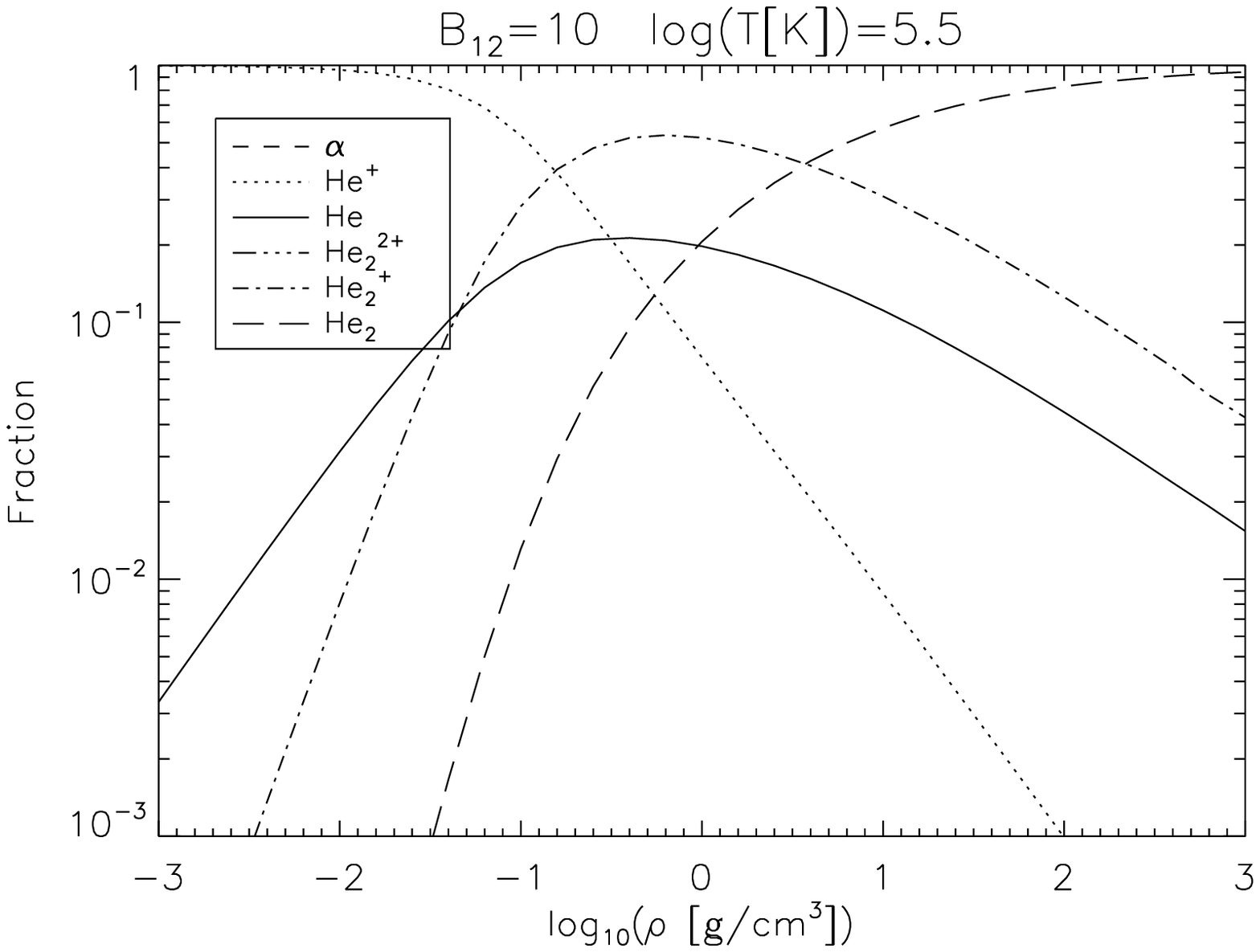}}
\resizebox{\hsize}{!}{\includegraphics{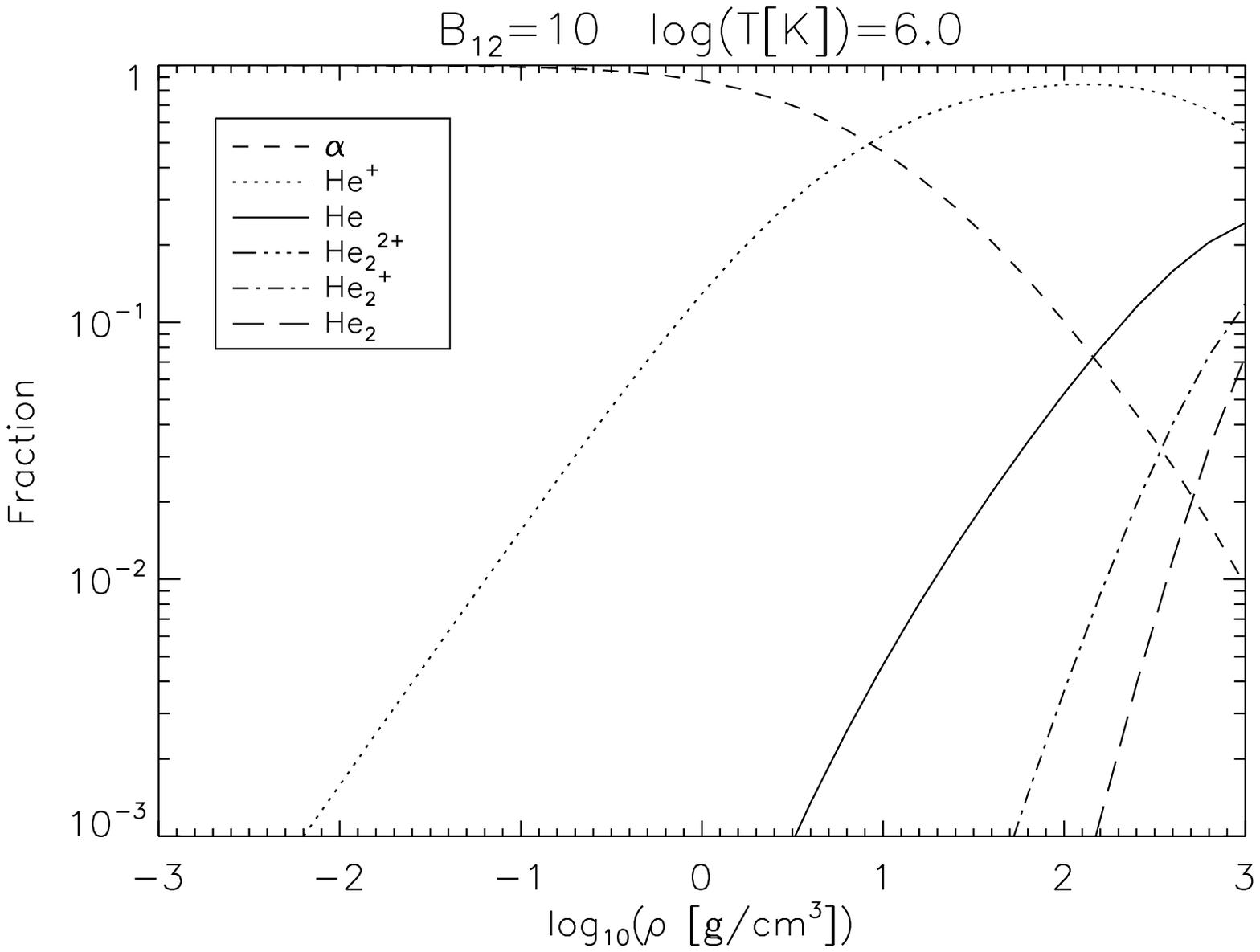}}
\caption{Ionization and dissociation balance of helium at  $T=10^{5.5}$ K (top) and $T=10^{6}$ K (bottom) at $B=10^{13}$ G. \label{fig_he13_rho}}
\end{figure}

\begin{figure} 
\resizebox{\hsize}{!}{\includegraphics{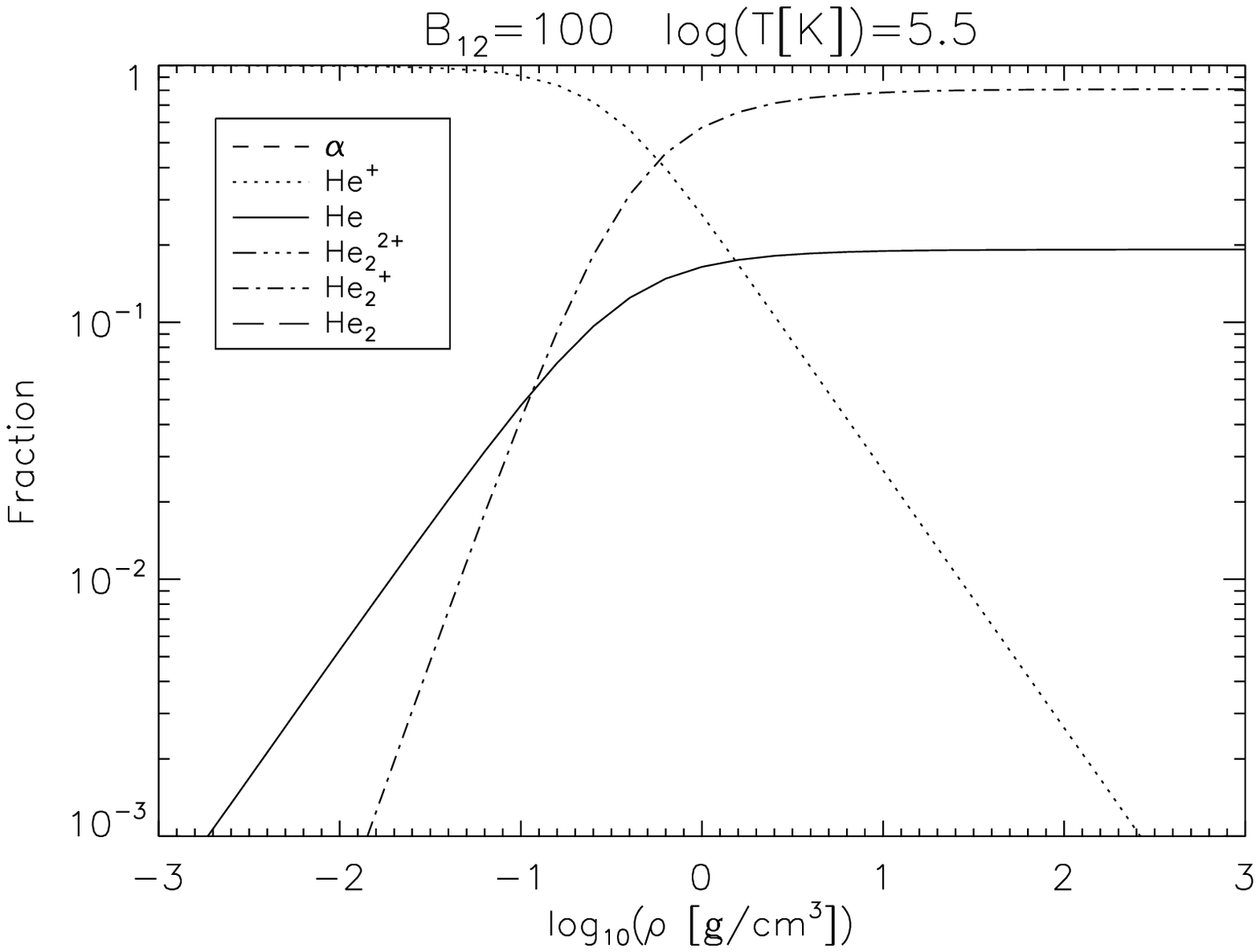}}
\resizebox{\hsize}{!}{\includegraphics{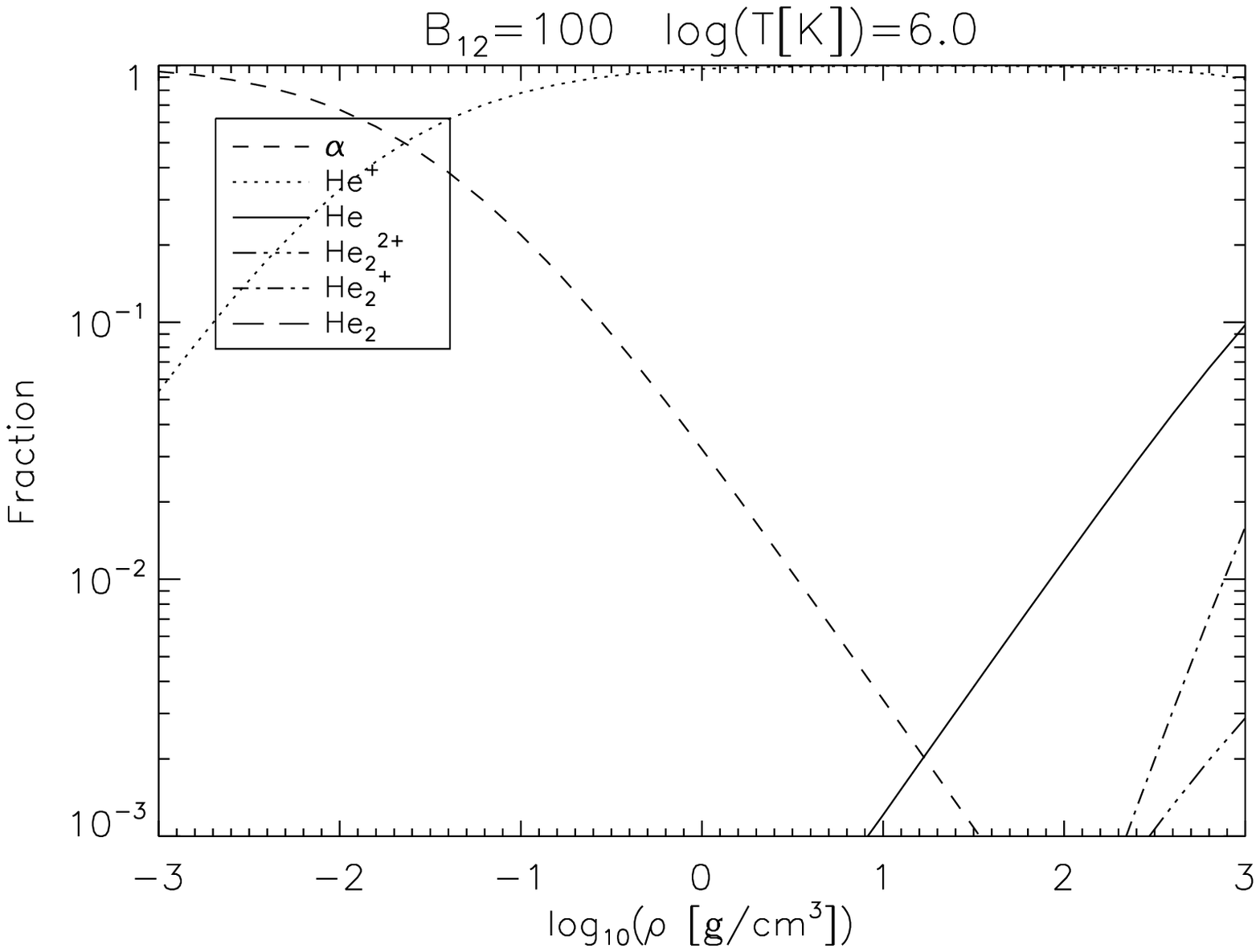}}
\caption{Ionization and dissociation balance of helium at  $T=10^{5.5}$ K (top) and $T=10^6$ K (bottom) at $B=10^{14}$ G. \label{fig_he14_rho}}
\end{figure}

\begin{figure} 
\resizebox{\hsize}{!}{\includegraphics{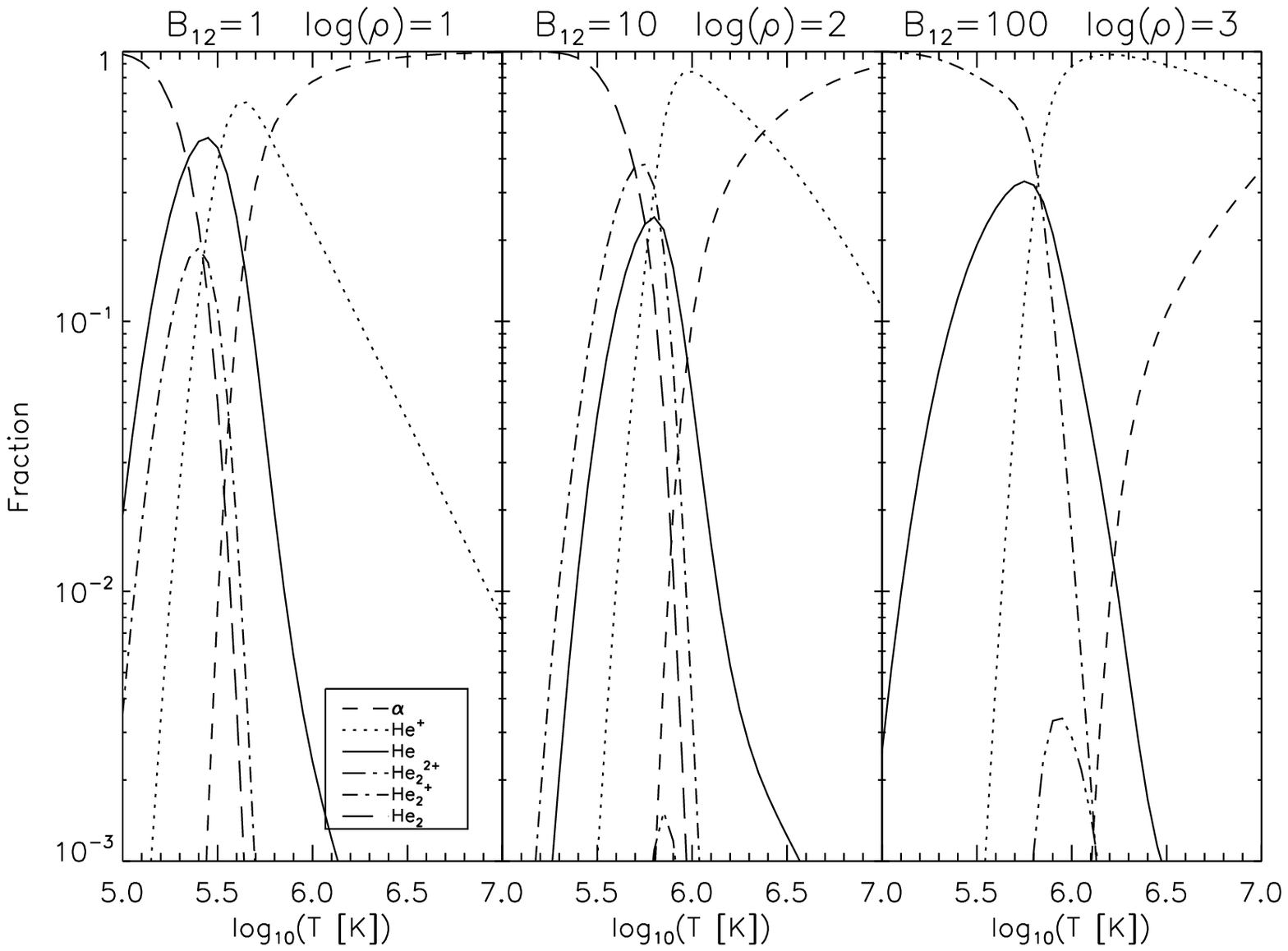}}
\caption{Temperature dependence of helium atomic and molecular
  fraction at $B=10^{12}$ G ($\rho=10$ g/cm$^3$), $B=10^{13}$ G ($\rho=10^2$  g/cm$^3$) and $B=10^{14}$ G ($\rho=10^3$ g/cm$^3$). \label{fig_temp}}
\end{figure}

Figure \ref{fig_he12_rho}, \ref{fig_he13_rho} and \ref{fig_he14_rho}
show the fraction of helium ions and molecular ions at $B=10^{12},
10^{13}$ and $10^{14}$ G. At $B=10^{12}$ G, He$_2^{3+}$ and
He$_2^{2+}$ are not present because they are not bound with respect to
their dissociated atoms and ions (table \ref{tab_dis_e}). In all the
cases, the He$_2^{3+}$ fraction is negligible because helium molecular
ions with more electrons have much larger binding energies. At
$B=10^{14}$ G, He$_2$ is not present because even the ground state
becomes auto-ionized to He$_2^{+}$ ion due to the finite nuclear mass
effect although He$_2$ remains bound with respect to two helium
atoms. Note that the ionization energy of He$_2$ at $10^{14}$ G is 643
eV (table \ref{tab_ion_e}) while the difference in the energy due to
the finite nuclear mass term between He$_2$ and He$_2^+$ is 945 eV. As
discussed in \S\ref{sec_nuc_mass}, our scheme using the zero
pseudomomentum becomes invalid at
$B \uax B_Q$ and in reality He$_2$ will have lower binding energy than
He$^+_2$.  Therefore, our results at $B=10^{14}$ G should be taken
with caution.  Figure~\ref{fig_temp} shows the temperature dependence
of the helium atomic and molecular fractions at different B-field
strengths. We fixed the plasma density to the typical density of the X-mode photosphere ($\sim10, 10^2$ and $10^3$ g/cm$^3$ for $B=10^{12}, 10^{13}$
and $10^{14}$ G \citep{lai97}). The transition from molecules to helium
atoms and ions takes place rapidly at $T\sim3\times10^5 $ K and
$\sim6\times10^5$ K for $B=10^{12}$ G and $10^{13}$ G.

\begin{figure} 
\resizebox{\hsize}{!}{\includegraphics{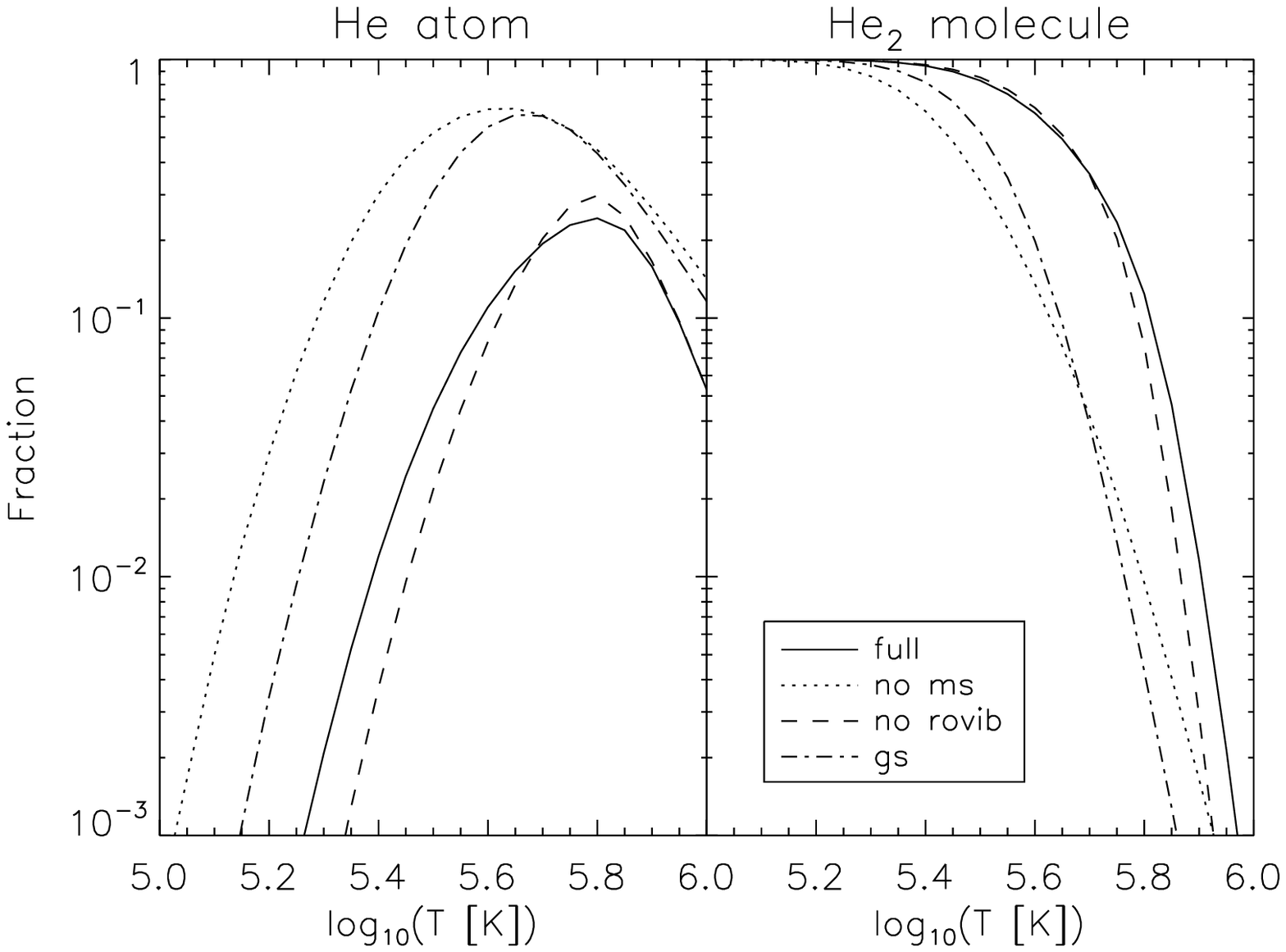}}
\caption{Fraction of helium atoms (left) and helium molecules (right)
  $\rho=10^2$ g/cm$^3$ at $B=10^{13}$ G. The solid lines show the
  results by taking into account all the physical effects discussed in
  this paper. The dotted and dashed lines ignored motional Stark
  effects and rotovibrational states respectively. The dot-dashed lines
  include only the ground states. \label{fig_he_comp}}
\end{figure}

In order to illustrate different physical effects, we show the
fraction of helium atoms and He$_2$ molecules as a function of
temperature in Figure~\ref{fig_he_comp}. When motional Stark effects
are ignored (dotted line), the molecular fraction is underestimated
because helium ions and atoms are subject to a larger motional Stark
field due to their smaller masses and binding energies. As expected,
molecules become more abundant when rotovibrational states are
included. The discrepancy from the results without rotovibrational
states (dashed line) increases toward higher temperature since more
rotovibrational states have larger statistical weight as their
excitation energies are an order of 100 eV at $B=10^{13}$ G (see
figure \ref{fig_he2_rovib}). When we take into account only the ground
states (dot-dashed line), the results are close to the case without
motional Stark effects because the ground states are less affected by
motional Stark effects than the excited states.

\subsection{He$_3$ molecule and larger helium molecular chains}

When He$_2$ is abundant, larger helium molecules such as He$_3$ may
become abundant. We roughly estimated the fraction of He$_3$ molecules
by neglecting the finite nuclear mass effects and zero-point
vibrational energy correction both of which will reduce the dissociation
energy.  Similarly to He$_2$, we computed the binding energy of He$_3$
molecules by Hartree-Fock calculation.  The dissociation energy of
He$_3\rightarrow$He$_2$+He is 289 and 1458 eV at $B=10^{13}$ and
$10^{14}$ G. Our results are close to those of the density functional
calculation by ML06 (384 and 1647 [eV] at $B=10^{13}$ and $10^{14}$
G). Note that the density function calculation overestimates binding
energies by $\sim10$\% compared with more accurate Hartree-Fock
calculation (ML06).  Assuming the internal degrees of freedom are
similar between He$_2$ and He$_3$ molecule and neglecting the bending
degree of freedom for He$_3$, we computed dissociation equilibrium
between He$_2$ and He$_3$ following \citet{lai97}.

\begin{figure} 
\resizebox{\hsize}{!}{\includegraphics{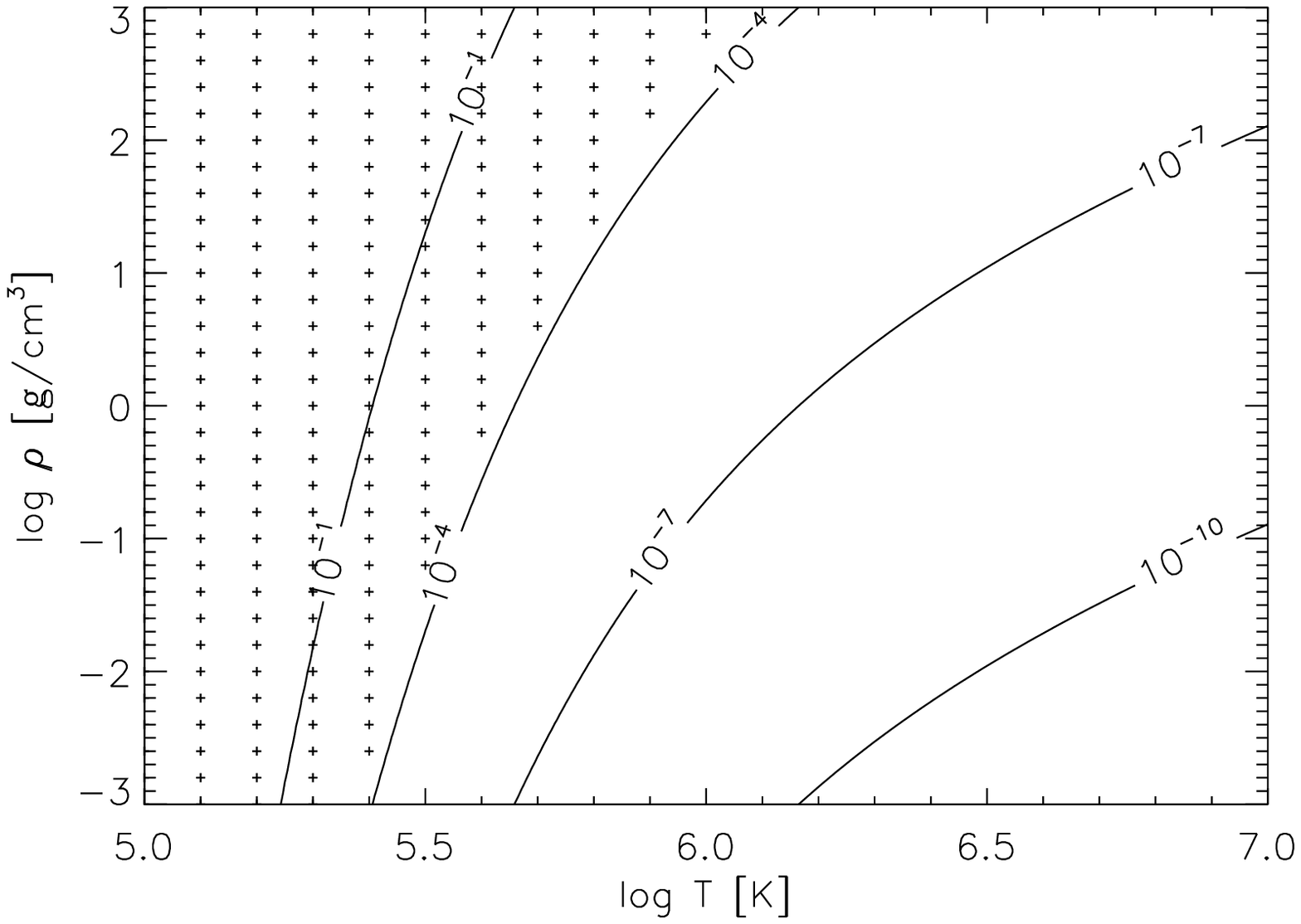}}
\caption{Ratio of He$_3$ with respect to He$_2$ at $B=10^{13}$ G
  (solid curve contours). The dotted
  area shows the $(\rho,T)$ regime where a fraction of diatomic helium
  molecular ions is larger than 10\%. 
  \label{fig_he3}}
\end{figure}

Figure~\ref{fig_he3} shows contours of the He$_3$ molecule fraction
with respect to He$_2$ molecule at $B=10^{13}$ G. In the dotted area,
the fraction of diatomic helium molecular ions is larger than 10\%. It
is seen that He$_3$ molecule will be abundant at $T\la3\times10^{5}$ K
at $B=10^{13}$ G. The results are consistent with the estimates of
molecular chain formation by \citet{lai01} and ML06. ML06 provided
binding energy of infinite He molecular chain as
$E_\infty\simeq1.25B^{0.38}_{13}$ keV at $B\ga10^{13}$ G. From this
fitting formula, the cohesive energy of helium molecular chains is
given as $E_{co}\sim0.36, 1.5$ and 5.1 keV at $B=10^{13}, 10^{14}$ and
$10^{15}$ G. When $kT\la 0.1E_{co}$, molecular chains are likely to be
formed \citep{lai01}; consequently, we expect helium molecular chains
to form at $T\la3\times10^5, 1\times10^6$ and $5\times10^6$ K and
$B\ga10^{13}, 10^{14}$ and $10^{15}$ G.

\section{Application} 

We investigated the ionization and dissociation balance of helium
atmospheres for two classes of INS whose X-ray spectra show absorption
features.      

\subsection{Radio-quiet neutron stars}

\begin{figure} 
\resizebox{\hsize}{!}{\includegraphics{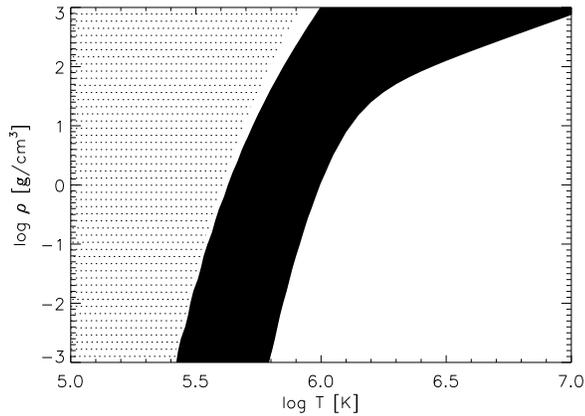}}
\caption{Ionization and dissociation balance of helium at
  $B=3\times10^{13}$ G.  The dotted area is where the fraction of
  helium molecules and molecular ions (including He$_2$, He$_2^+$,
  He$_2^{2+}$ and He$_2^{3+}$) becomes more than 50\%.  The black area
  is where the fraction of He$^+$ becomes more than 50\%. The white
  region on the right is where the fraction of bare helium ions becomes
  more than 50\%.  In the narrow white region betweeen the dotted and
  black area, atomic helium is somewhat abundant, so neither molecular
  nor singly ionized helium dominates. \label{fig_he30}}
\end{figure}

A class of INS called radio-quiet neutron stars (RQNS) is
characterized by their X-ray thermal spectra with $T\la10^6$ K and
spin-down dipole B-field strength $B\ga10^{13}$ G. A single or
multiple absorption features have been detected from six radio-quiet
NS \citep{haberl03, haberl04, vankerkwijk04}. Although the
interpretation of these features is still in debate, helium is
certainly one of the candidates for the surface composition of RQNS
\citep{vankerkwijk06}. Timing analysis suggests that a couple of RQNS
have dipole magnetic field strengths in the range of
$B=10^{13}$--$10^{14}$~G \citep{vankerkwijk05}. Therefore, we
investigated the ionization/dissociation balance of a helium atmosphere at
$B=3\times10^{13}$ G. Figure~\ref{fig_he30} shows the contours of He$^+$
and He molecular fractions at $3\times10^{13}$ G. He$^+$ is dominant at
$T\sim10^6$ K and molecules become largely populated at
$T\la5\times10^5$ K. 

Suppose all the RQNS have helium atmospheres on the
surface. RQNS with higher temperatures ($T\sim10^6$ K) such as
RXJ0720.4-3125 and RXJ1605.3+3249 will have He$^+$ ions predominantly
with a small fraction of He atoms. Indeed, several bound-bound
transition lines of He$^+$ ion have energies at the observed
absorption line location at $B\ga10^{13}$ G \citep{pavlov05}. On the
other hand, RQNS with lower temperatures ($T\sim5\times10^5$ K) such as
RXJ0420.0-5022 may have some fraction of helium molecules, although
the He$^+$ ion is still predominant at low densities where absorption
lines are likely formed.

\subsection{1E1207.4-5209} 

1E1207.4-5209 is a hot isolated NS \footnote{Recent timing analysis
  suggests that 1E1207 is in a binary system \citep{woods06}. However,
  it is unlikely that 1E1207 is an accreting NS.} with age
$\sim7\times10^3$~yrs. The fitted blackbody temperature is
$\sim2\times10^6$~K \citep{mori05}. Presence of a non-hydrogenic
atmosphere has been suggested since 1E1207 shows multiple absorption
features at higher energies than hydrogen atmosphere models predicted
\citep{sanwal02, hailey02, pavlov05, mori06}. \citet{sanwal02}
interpreted the observed features as bound-bound transition lines of
a He$^+$ ion at $B=2\times10^{14}$ G. Also, \citet{turbiner05} suggested
that He$^{3+}_2$ molecular ion may be responsible for one of the
absorption features observed in 1E1207 at $B\sim4.4\times10^{13}$ G.

\begin{figure} 
\resizebox{\hsize}{!}{\includegraphics{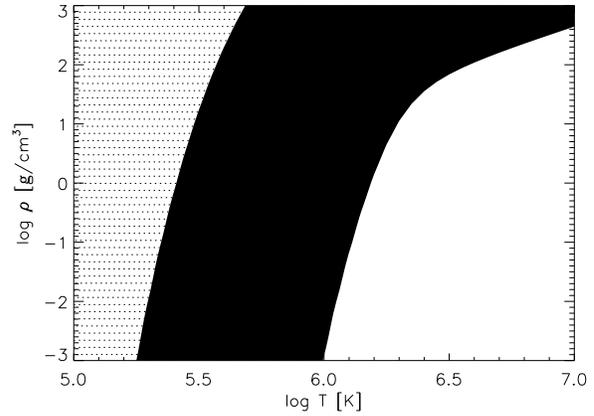}}
\resizebox{\hsize}{!}{\includegraphics{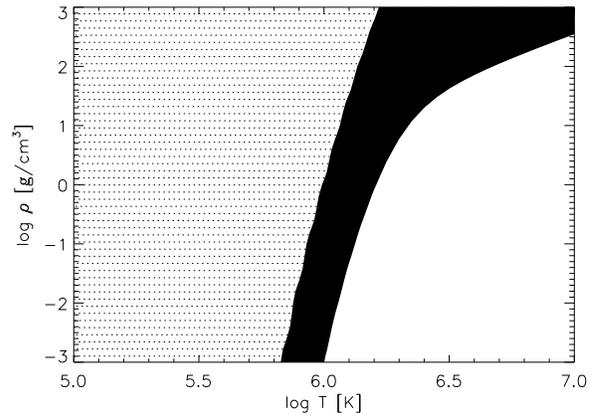}}
\caption{Ionization and dissociation balance of helium at
  $B=2\times10^{14}$ G. The top and bottom figure show the results
  with and without the finite nuclear mass effects respectively. In
  the dotted region the molecular and molecular ion fraction exceeds
  one half.  In the black region singly ionized atomic helium
  dominates, and the white region demarcates fully ionized
  helium.\label{fig_he200}}
\end{figure}

However, our study shows that the fraction of He$^{3+}_2$ molecular ions
is negligible at any B-field and temperature because more neutral
molecular ions have significantly larger binding energies.  We also
investigated ionization/dissociation balance of helium atmospheres at
$2\times10^{14}$ G (figure \ref{fig_he200}).  At $B \uax B_{Q}$, our
scheme of treating finite nuclear mass effects becomes progressively
inaccurate. As a result, the ionization energy of helium atoms becomes
significantly smaller and the He$_2$ molecule becomes auto-ionized due
to the finite nuclear mass correction discussed in
\S\ref{sec_nuc_mass}. Therefore, the molecular fraction will be
underestimated in this case (the left panel in figure
\ref{fig_he200}).  On the other hand, when we ignore the finite
nuclear mass effects (the right panel in figure \ref{fig_he200}), the
molecular fraction is likely overestimated. We expect that a realistic
ionic and molecular fraction will be somewhere between the two
cases. It is premature to conclude whether He$^+$ is dominant for the
case of 1E1207 since we do not have a self-consistent temperature and
density profile. The study of ML06 and \citet{lai01} also suggests the
critical temperature below which helium molecule chains form is
$\sim2\times10^6$ K at $B=2\times10^{14}$ G. This is close to the
blackbody temperature of 1E1207. Further detailed studies are
necessary to conclude the composition of helium atmospheres at
$B=2\times10^{14}$ G.

\section{Discussion}

We have examined the ionization-dissociation balance of the helium
atmospheres of strongly magnetized neutron stars.  As the
observational data on isolated neutron stars has improved over the
past decade \citep{sanwal02, hailey02, vankerkwijk06}, both hydrogen and
iron atmospheres have lost favour, and atmospheres composed of
other elements have been invoked to interpret both the continuum
and spectral properties of neutron stars.  On the other hand the
theoretical investigations of \citet{chang04_1} and
\citet{chang04_2} have shown that diffusive nuclear burning may
quickly deplete the hydrogen by so the NS surface may be composed of
helium or heavier elements.

The results found here indicate several avenues for future
work.  Although the treatment of finite nuclear mass effects is problematic in
the strong field limit, only by including this accurately can we know
the physical state of matter in super-critical magnetic fields --
solid, liquid or gas.  The state can have significant effects of the
outgoing radiation from these objects.  It is also important to repeat
a similar calculation to that presented here for molecular chains,
including all of the relevant degrees of freedom of the chains, {\em
  i.e.} including the bending modes which were neglected here.
Regardless, the high abundance of molecules under the conditions of
observed neutron star atmospheres is bound to spur additional research
into the statistical mechanics of highly magnetized molecules and
polymers.

At the temperatures and densities of neutron star atmospheres the
rotovibrational excitations of helium molecules are populated.
Including these excitations increases the expected abundance of
molecules by up to two orders of magnitude relative to calculations
that ignore the internal states of the molecule.  
Ionization, dissociation and electric excitation energies of helium
molecules are larger than 100 eV at $B\uax10^{13}$ G. On the other
hand, rotovibrational excitation energies are in the range of 10--100
eV at $B=10^{12}$--$10^{14}$ G. If helium molecules are abundannt,
their spectroscopic signatures may be detected in the optical, UV and X-ray band.   
If helium comprises the atmospheres of isolated neutron stars, clearly it is crucial to
understand the structure of helium molecules and molecular chains in
order to interpret the spectra from neutron stars.

\section*{acknowledgments}
  We would like to thank the anonymous referee for comments that
  helped to improve the paper. KM acknowledges useful discussions with
  George Pavlov and Marten van Kerkwijk.   
  This work was supported in part by a Discovery Grant from NSERC
  (JSH).  This work made use of NASA's Astrophysics Data System.  The
  authors were visitors at the Pacific Institute of Theoretical
  Physics during the nascent stages of this research.



\begin{thebibliography}{41}
\expandafter\ifx\csname natexlab\endcsname\relax\def\natexlab#1{#1}\fi

\bibitem[{{Alcock} \& {Illarionov}(1980)}]{alcock80}
{Alcock}, C. \& {Illarionov}, A. 1980, \apj, 235, 534

\bibitem[{{Bezchastnov} {et~al.}(1998){Bezchastnov}, {Pavlov}, \&
  {Ventura}}]{bezchastnov98}
{Bezchastnov}, V.~G., {Pavlov}, G.~G., \& {Ventura}, J. 1998, \pra, 58, 180

\bibitem[{{Chang} {et~al.}(2004){Chang}, {Arras}, \& {Bildsten}}]{chang04_1}
{Chang}, P., {Arras}, P., \& {Bildsten}, L. 2004, \apjl, 616, L147

\bibitem[{{Chang} \& {Bildsten}(2004)}]{chang04_2}
{Chang}, P. \& {Bildsten}, L. 2004, \apj, 605, 830

\bibitem[{{Demeur} {et~al.}(1994){Demeur}, {Heenen}, \& {Godefroid}}]{demeur94}
{Demeur}, M., {Heenen}, P.~., \& {Godefroid}, M. 1994, \pra, 49, 176

\bibitem[{{Gnedin} {et~al.}(1974){Gnedin}, {Pavlov}, \& {Tsygan}}]{gnedin74}
{Gnedin}, Y.~N., {Pavlov}, G.~G., \& {Tsygan}, A.~I. 1974, Soviet Physics JETP,
  39, 201

\bibitem[{{Haberl} {et~al.}(2003){Haberl}, {Schwope}, {Hambaryan}, {Hasinger},
  \& {Motch}}]{haberl03}
{Haberl}, F., {Schwope}, A.~D., {Hambaryan}, V., {Hasinger}, G., \& {Motch}, C.
  2003, \aap, 403, L19

\bibitem[{{Haberl} {et~al.}(2004){Haberl}, {Zavlin}, {Tr{\" u}mper}, \&
  {Burwitz}}]{haberl04}
{Haberl}, F., {Zavlin}, V.~E., {Tr{\" u}mper}, J., \& {Burwitz}, V. 2004, \aap,
  419, 1077

\bibitem[{{Hailey} \& {Mori}(2002)}]{hailey02}
{Hailey}, C.~J. \& {Mori}, K. 2002, \apjl, 578, L133

\bibitem[{{Herold} {et~al.}(1981){Herold}, {Ruder}, \& {Wunner}}]{herold81}
{Herold}, H., {Ruder}, H., \& {Wunner}, G. 1981, Journal of Physics B Atomic
  Molecular Physics, 14, 751

\bibitem[{{Kaplan} \& {van Kerkwijk}(2005)}]{vankerkwijk05}
{Kaplan}, D.~L. \& {van Kerkwijk}, M.~H. 2005, \apj, 635, L65

\bibitem[{{Khersonskii}(1984)}]{khersonskii84}
{Khersonskii}, V.~K. 1984, AP\&SS, 98, 255

\bibitem[{{Khersonskii}(1985)}]{khersonskii85}
---. 1985, Ap\&SS, 117, 47

\bibitem[{{Lai}(2001)}]{lai01}
{Lai}, D. 2001, Rev. Mod. Phys., 73, 629

\bibitem[{{Lai} \& {Salpeter}(1995)}]{lai95}
{Lai}, D. \& {Salpeter}, E.~E. 1995, \pra, 52, 2611

\bibitem[{{Lai} \& {Salpeter}(1996)}]{lai96}
---. 1996, \pra, 53, 152

\bibitem[{{Lai} \& {Salpeter}(1997)}]{lai97}
---. 1997, \apj, 491, 270

\bibitem[{{Lai} {et~al.}(1992){Lai}, {Salpeter}, \& {Shapiro}}]{lai92}
{Lai}, D., {Salpeter}, E.~E., \& {Shapiro}, S.~L. 1992, \pra, 45, 4832

\bibitem[{{Medin} \& {Lai}(2006{\natexlab{a}})}]{medin06_1}
{Medin}, Z. \& {Lai}, D. 2006{\natexlab{a}}, submitted to Phys. Rev. A
  (astro-ph/0607166)

\bibitem[{{Medin} \& {Lai}(2006{\natexlab{b}})}]{medin06_2}
---. 2006{\natexlab{b}}, submitted to Phys. Rev. A (astro-ph/0607277)

\bibitem[{{Mori} {et~al.}(2005){Mori}, {Chonko}, \& {Hailey}}]{mori05}
{Mori}, K., {Chonko}, J.~C., \& {Hailey}, C.~J. 2005, \apj, 631, 1082

\bibitem[{{Mori} \& {Hailey}(2002)}]{mori02}
{Mori}, K. \& {Hailey}, C.~J. 2002, \apj, 564, 914

\bibitem[{{Mori} \& {Hailey}(2006)}]{mori06}
---. 2006, \apj, 648, 1139

\bibitem[{{Morse}(1929)}]{morse29}
{Morse}, P.~M. 1929, Physical Review, 34, 57

\bibitem[{{Neuhauser} {et~al.}(1987){Neuhauser}, {Koonin}, \&
  {Langanke}}]{neuhauser87}
{Neuhauser}, D., {Koonin}, S.~E., \& {Langanke}, K. 1987, \pra, 36, 4163

\bibitem[{{Pavlov} \& {Bezchastnov}(2005)}]{pavlov05}
{Pavlov}, G.~G. \& {Bezchastnov}, V.~G. 2005, \apjl, 635, L61

\bibitem[{{Pavlov} \& {Meszaros}(1993)}]{pavlov93}
{Pavlov}, G.~G. \& {Meszaros}, P. 1993, \apj, 416, 752

\bibitem[{{Potekhin}(1994)}]{potekhin94}
{Potekhin}, A.~Y. 1994, J. Phys. B, 27, 1073

\bibitem[{{Potekhin}(1998)}]{potekhin98-2}
---. 1998, J. Phys. B, 31, 49

\bibitem[{{Potekhin} {et~al.}(2002){Potekhin}, {Chabrier}, \&
  {Gilles}}]{potekhin02}
{Potekhin}, A.~Y., {Chabrier}, G., \& {Gilles}, D. 2002, Phys. Rev. E, 65,
  36412

\bibitem[{{Romani}(1987)}]{romani87}
{Romani}, R.~W. 1987, \apj, 313, 718

\bibitem[{{Sanwal} {et~al.}(2002){Sanwal}, {Pavlov}, {Zavlin}, \&
  {Teter}}]{sanwal02}
{Sanwal}, D., {Pavlov}, G.~G., {Zavlin}, V.~E., \& {Teter}, M.~A. 2002, \apjl,
  574, L61

\bibitem[{{Turbiner}(2005)}]{turbiner05}
{Turbiner}, A.~V. 2005, preprint (astro-ph/0506677)

\bibitem[{{Turbiner} \& {Guevara}(2006)}]{turbiner06_3}
{Turbiner}, A.~V. \& {Guevara}, N.~L. 2006, preprint (astro-ph/0610928)

\bibitem[{{Turbiner} \& {L{\' o}pez Vieyra}(2004)}]{turbiner04_2}
{Turbiner}, A.~V. \& {L{\' o}pez Vieyra}, J.~C. 2004, preprint
  (astro-ph/0412399)

\bibitem[{{Turbiner} \& {L{\'o}pez Vieyra}(2003)}]{turbiner03}
{Turbiner}, A.~V. \& {L{\'o}pez Vieyra}, J.~C. 2003, \pra, 68, 012504

\bibitem[{{van Kerkwijk} \& {Kaplan}(2006)}]{vankerkwijk06}
{van Kerkwijk}, M.~H. \& {Kaplan}, D.~L. 2006, to appear in Astrophysics and
  Space Science, in the proceedings of "Isolated Neutron Stars: from the
  Interior to the Surface", edited by D. Page, R. Turolla and S. Zane, preprint
  (astro-ph/0607320)

\bibitem[{{van Kerkwijk} {et~al.}(2004){van Kerkwijk}, {Kaplan}, {Durant},
  {Kulkarni}, \& {Paerels}}]{vankerkwijk04}
{van Kerkwijk}, M.~H., {Kaplan}, D.~L., {Durant}, M., {Kulkarni}, S.~R., \&
  {Paerels}, F. 2004, \apj, 608, 432

\bibitem[{{Vincke} {et~al.}(1992){Vincke}, {LeDourneuf}, \& {Baye}}]{vincke92}
{Vincke}, M., {LeDourneuf}, M., \& {Baye}, D. 1992, Journal of Physics B Atomic
  Molecular Physics, 25, 2787

\bibitem[{{Woods} {et~al.}(2006){Woods}, {Zavlin}, \& {Pavlov}}]{woods06}
{Woods}, P.~M., {Zavlin}, V.~E., \& {Pavlov}, G.~G. 2006, to appear in
  Astrophysics and Space Science, in the proceedings of "Isolated Neutron
  Stars: from the Interior to the Surface", edited by D. Page, R. Turolla and
  S. Zane, preprint (astro-ph/0608483)

\bibitem[{{Wunner} {et~al.}(1981){Wunner}, {Ruder}, \& {Herold}}]{wunner81}
{Wunner}, G., {Ruder}, H., \& {Herold}, H. 1981, \apj, 247, 374

\end{thebibliography}

\label{lastpage}

\end{document}